\documentclass[12pt]{article}
\usepackage{amsmath,amssymb}
\usepackage[dvips]{graphicx}
\usepackage{booktabs}
\renewcommand{\theequation}{\thesection.\arabic{equation}}
\renewcommand{\thefootnote}{\fnsymbol{footnote}}
\newcommand{\beq}{\begin{equation}}
\newcommand{\eeq}{\end{equation}}
\newcommand{\bea}{\begin{eqnarray}}
\newcommand{\ena}{\end{eqnarray}}
\newcommand{\ba}{\begin{align}}
\newcommand{\ea}{\end{align}}
\newcommand{\vs}[1]{\vspace{#1 mm}}

\newcommand{\uda}{\nearrow \kern-1em \searrow}
\makeatletter
\def\eqnarray{%
 \stepcounter{equation}%
 \let\@currentlabel=\theequation
 \global\@eqnswtrue
 \global\@eqcnt\z@
 \tabskip\@centering
 \let\\=\@eqncr
 $$\halign to \displaywidth\bgroup\@eqnsel\hskip\@centering
 $\displaystyle\tabskip\z@{##}$&\global\@eqcnt\@ne
 \hfil$\displaystyle{{}##{}}$\hfil
 &\global\@eqcnt\tw@$\displaystyle\tabskip\z@{##}$\hfil
 \tabskip\@centering&\llap{##}\tabskip\z@\cr}
\makeatother

\def\cal{\mathcal}

\def\nonum{\nonumber}
\begin{document}
\begin{titlepage}
\setcounter{page}{0}
\begin{flushright}
EPHOU 10-005\\
\today\\
\end{flushright}

\vs{6}
\begin{center}
{\Large \textbf{
Open mirror symmetry for Pfaffian Calabi-Yau $3$-folds}}

\vs{6}
\textbf{Masahide Shimizu \ and \ Hisao Suzuki}\\
{\small \em Department of Physics, 
Hokkaido University, \\Sapporo 060-0810, Japan } \\
\end{center}
\vs{6}

\centerline{{\bf{Abstract}}}
 We investigate the open mirror symmetry of certain non-complete intersection Calabi-Yau $3$-folds, 
called pfaffian Calabi-Yau. 
 We predict the number of disk invariants of some examples by using the direct integration method proposed recently and the open mirror symmetry. 
We treat several pfaffian Calabi-Yau $3$-folds and branes with two discrete vacua. 
Some models have two special points in its moduli space, around both of which we can consider different A-model mirror partners.  We compute disc invariants for both cases.
This study  is the first application of the open mirror symmetry to the compact non-complete intersections in toric variety. 

\end{titlepage}
\newpage

\renewcommand{\thefootnote}{\arabic{footnote}}
\setcounter{footnote}{0}

\tableofcontents
\section{Introduction}

 Recently, the research on open mirror symmetry for {\it compact} Calabi-Yau manifolds has been progressing remarkably. 
The first important step was achieved by Walcher, 
who showed that the Picard-Fuchs differential equations are modified by the D-branes \cite{Wa1}. Then the exact computation in the B-model side was done by calculating the inhomogeneous term of Picard-Fuchs equation \cite{MW}. 
The enumerative predictions of disk invariants were confirmed mathematically by the A-model localization computation \cite{PaSoWa}. 
The extension of the holomorphic anomaly equation \cite{BCOV1,BCOV2} to the open string sector was also developed \cite{Wa2,Wa3}. 
After these works, several important analyses and techniques have been proposed in many papers. 

One of difficulties comes from the fact that, for compact Calabi-Yaus, the open moduli (D-brane moduli) are obstructed in general and take discrete values. 
In \cite{JS1}, the authors discussed the so-called {\it off-shell} idea, 
by virtually introducing the continuous open moduli and 
by replacing the curve (D-brane) by the divisor. 
Then, the discrete {\it on-shell} situation is realized by taking critical values with respect to the continuous open moduli. 
By this prescription, we can apply the ordinary 
method developed in non-compact case, namely $\mathcal{N}=1$ special geometry \cite{LMW1,LMW2}, to the compact case. 
After this work, related to this formalism, 
the application of the toric method to compact case was proposed in \cite{AHMM}, 
along with the original application of toric method to open mirror symmetry for non-compact case \cite{AV}. 
The direct integration of period, 
which is similar to the method in the classical work by Candelas et al. \cite{CDGP}, was also presented in \cite{JS2}. 
For several other developments, see \cite{KW1,KS1,GHKK,Walcher,AB,worldsheet} including our work \cite{FNSS}. 
It is worth noting that recently the relation of the open string on compact Calabi-Yau $3$-fold to the closed string on non-compact Calabi-Yau $4$-fold (and F-theory on the same CY$4$) were discussed in \cite{AB} in association with \cite{mayr}. 
Many different geometries, which lead to the same $4$-dimensional $\mathcal{N}=1$ theory, 
are related by the duality chain \cite{AHJMMS,GHKK2,GHKK3,JMW}.

The study of open mirror symmetry for compact Calabi-Yau is becoming very active, 
but the analyzed models are limited to the ones given by hypersurfaces/complete intersections with a few moduli\footnote{
While this paper was in preparation for submission, a related work appeared \cite{GKZ} and the analysis of open mirror symmetry on compact Calabi-Yau hypersurfaces with $2$- and $3$-moduli were carried out by a very systematic toric and GKZ-system approach. 
}, 
in toric varieties. 
However, there are many examples of non-complete intersection Calabi-Yau manifolds. 
Actually, there are useful lists of Calabi-Yaus in \cite{data,table}. 
In the database \cite{data}, a very large number of Calabi-Yau $3$-folds with one moduli are listed and studied by analyzing the Picard-Fuchs operators numerically. 
Only a small portion of the models can be expressed as complete intersections in weighted projective spaces or toric varieties. 
Most of the models listed in \cite{data} have not been expressed as hypersurfaces or complete intersections so far. 
It is suggested that the Calabi-Yau manifolds we know are not sufficient to find the string vacuum which may describe our world. 
So it is important to proceed to analyze the open mirror symmetry of the models which are not expressed as complete intersections.

In this paper, we treat certain non-complete intersection Calabi-Yau manifolds, so-called {\it pfaffian} Calabi-Yaus. 
The (closed) mirror phenomena of this type of Calabi-Yau manifolds were first discussed in \cite{Rodland}, and further investigations were done in \cite{HoKo, Kanazawa}. 
This type of Calabi-Yau manifolds have very interesting properties in the sense that 
some of them have two special points in their moduli spaces, around both of which we can consider the mirror phenomena. 
The corresponding Calabi-Yau varieties in the A-model side are different ones\footnote{
It is worth noting that these two Calabi-Yau varieties cannot be birationally equivalent but surprisingly their derived categories of coherent sheaf are equivalent. 
} 
and thus two different Calabi-Yau varieties are governed by the single moduli space in the mirror B-side. 
We can obtain enumerative predictions of certain geometric invariants by mirror symmetry at both points. 
In the present paper, we study the open mirror symmetry of this type of Calabi-Yau manifolds by our method recently proposed in \cite{FNSS}. 

This work is important from the two different viewpoints. 
One is the physical viewpoint and the other is the geometric one. 
Physically, 
we compute the superpotential of $4$-dimensional $\mathcal{N}=1$ effective field theory, 
which is obtained as the type II string theory compactified on Calabi-Yau with SUSY brane. 
Since Calabi-Yaus treated in this paper were not yet analyzed in the context of open mirror symmetry, 
it may be possible that the obtained effective superpotentials are useful for constructing realistic phenomenological/cosmological models. 
Geometrically, 
we would like to emphasize that this is the first results of analyzing the open mirror symmetry and counting the holomorphic discs, on compact non-complete intersection Calabi-Yau manifolds. 

Finally we note that we do not analyze at all the $A$-model side geometries in this paper. 
At present it is difficult to analyze the $A$-model side, 
especially for pfaffian case, 
since we cannot find how to construct the special Lagrangian submanifold. 
We also would like to note that, although we do not discuss the off-shell situation in this paper, our method is also applicable to the off-shell analysis as shown in \cite{FNSS}. 

The organization of this paper is as follows. 
In section \ref{background}, 
we will review some basic backgrounds for both physics and mathematics, 
including several definitions and notations needed in later sections. 
We will also introduce several models and their geometric data. 
In section \ref{formal}, 
we will discuss the formal aspects of the direct integration method, especially focusing on the application to the pfaffian Calabi-Yau manifolds. 
In section \ref{main}, 
we shall carry out the concrete computations of the direct integration for several models in detail. 
We will consider both large and small moduli limits. 
The resulting superpotentials will be interpreted as the A-model quantity
and the number of disk invariants will be predicted. 
In section \ref{one-loop}, we will discuss the one-loop amplitudes for consistency check, by the use of the extended holomorphic anomaly equation. 
Section \ref{conclusion} contains conclusions and discussions. 
Finally, in appendix \ref{invariants}, we will list tables of real BPS invariants.

\section{Open mirror symmetry and Pfaffian Calabi-Yau manifold}
\label{background}
In this section we review some physical and mathematical backgrounds needed in later sections. 

\subsection{Open mirror symmetry and D-brane superpotential}
We start with the following set up: 
type II string theory on $\mathbb{R}^{1,3}\times \text{Calabi-Yau}$ with 
a D-brane which is located in the entire $\mathbb{R}^{1,3}$ and wrapped on a SUSY cycle in Calabi-Yau. 
By the SUSY cycle, we mean a special Lagrangian submanifold for type IIA (A-model), and a holomorphic submanifold for type IIB (B-model) \cite{BBS,OOY}. 
Then we obtain the $4$-dimensional $\mathcal{N}=1$ supersymmetric gauge theory. 
The background D-brane wrapping around SUSY cycle contributes to the effective superpotential of $4$-dimensional theory \cite{OV}. 
What we would like to compute is this effective superpotential induced by D-branes. 

We consider the tree level of the topological B-model. 
This is described by the so-called special geometry, which is related to the deformation theory of the complex structure (Hodge structure) of Calabi-Yau $3$-fold. 
The important quantities are the periods which are defined as the integrals of the holomorphic $3$-form over several $3$-cycles. 
They are holomorphic with respect to the complex structure moduli (closed string moduli) and solutions of the Picard-Fuchs differential equation. 

The open string tree-level data is governed by the extended Picard-Fuchs equation, which generally depends on both open and closed moduli. 
This framework is described by the so-called $\mathcal{N}=1$ special geometry \cite{LMW1,LMW2}, and related to the mathematical theory of mixed Hodge structure. 
The relative periods, which are defined as the integrals of the holomorphic $3$-form over $3$-chains, play important roles. 
They contain information of the deformation theory of bulk Calabi-Yau and D-brane (holomorphic submanifold, typically holomorphic curve) in it. 
The D-brane superpotential encodes obstructions of deformations of this holomorphic curve \cite{KKLM}. 

Now we shall focus on the {\it compact} Calabi-Yau case. 
Due to the fact that, in general, the open moduli take discrete values, the net result turns out to be the inhomogeneous modification of the ordinary Picard-Fuchs equation \cite{Wa1,MW}, i.e. 
\beq
\mathcal{L}_{\text{PF}}\varpi_i=f(z), \label{mpf}
\eeq
where $f(z)$ is a certain function of closed moduli $z$. 
One of solutions of this inhomogeneous Picard-Fuchs equation is a superpotential (or a domainwall tension, see below). 
Let us recall the typical situation in the B-model side: 
D5-brane which is wrapped on two holomorphic curves $C_\pm$ in Calabi-Yau 
and located in the entire $\mathbb{R}^{1,3}$. 
Moreover, the curves $C_+$ and $C_-$ are homologous each other and there exists a $3$-chain $\Gamma$ which interpolates between them (namely, $\partial\Gamma=[C_+-C_-]$). 
Physically, $C_\pm$ correspond to two supersymmetric vacua of $4$-dimensional $\mathcal{N}=1$ supersymmetric theory and we can find the BPS domainwall which wraps $\Gamma$. 
The tension of this BPS domainwall is equal to the difference of the superpotentials of $C_\pm$, and is given by the integral 
of the holomorphic $3$-form over the $3$-chain $\Gamma$ \cite{AV,Witten}
\beq
\int_{\Gamma(C_+-C_-)}\Omega^{3,0}. 
\eeq
This formula can be obtained by the dimensional reduction of the (topological) D-brane world-volume action (i.e., holomorphic Chern-Simons action \cite{hCS}) on the internal part of the D5-brane (i.e., the curve in Calabi-Yau). 
This contributes to the effective superpotential of $4$-dimensional theory exactly \cite{OV}. 
In mathematical language, this is a so-called normal function \cite{MW} (for mathematical ingredients, see e.g. \cite{Griffiths,Griffiths2,Green}, and the references therein). 
This domainwall tension can be obtained as one of the solutions of the inhomogeneous version of the Picard-Fuchs equation \eqref{mpf}. 
The method used to compute the inhomogeneous term was proposed in \cite{MW,LLY}. 
In cases of $\mathbb{Z}_2$-brane in hypersurface/complete intersection, the inhomogeneous term in \eqref{mpf} takes the form $f(z) \sim z^{1/2}$ \footnote{
Interestingly, this simple form is modified for pfaffian Calabi-Yaus treated in this paper, as discussed in later sections. 
}. 
Although this method is totally rigorous and systematic, it is generally very tedious task. 
Therefore, another method to compute on-shell D-brane superpotential was proposed in \cite{FNSS}, which will be explained in detail in section \ref{formal}. 

The so-called {\it off-shell} method has also been developed, 
by introducing the continuous open moduli and replacing the curve by a divisor \cite{JS1}. 
By this picture, we can apply the framework of $\mathcal{N}=1$ special geometry \cite{LMW1,LMW2} to the compact case \cite{JS1}. 
Toric method \cite{AV} was also applied to the compact case \cite{AHMM}. 
The picture established in \cite{JS1} is that the on-shell superpotential is realized by restricting the off-shell superpotential to the critical values with respect to brane moduli. 
For example, in $\mathbb{Z}_2$-vacua case, 
\beq
\mathcal{W}(z,\pm)=\mathcal{W}(z,u)\Big|_{\partial_u\mathcal{W}=0}, 
\eeq
where $z$ is a bulk moduli and $u$ is a brane moduli. 
For several other important developments, see \cite{JS2,AB,worldsheet,GKZ,LLY}. 
Although the off-shell analysis is very important and has a rich structure, 
we will not make such an analysis in the present paper. 

What about the A-type D-brane superpotential? 
The A-type D-brane superpotential at the tree level contains non-perturbative effects, 
namely, the effects of disk instantons. 
They are contributions from holomorphic maps of a disk worldsheet into $X$, with a boundary on $L$. 
The superpotential can be viewed as a generating function of disk invariants (or BPS invariants). 
At present, the A-model computations of disc instantons (namely, the localization technique) for compact Calabi-Yau manifolds are limited to complete intersections in projected spaces. 
The direct computation of the A-brane superpotential is difficult but the assumption of open mirror symmetry enables us to compute this quantity exactly by reinterpreting the B-brane superpotential.

Mirror symmetry relates Calabi-Yau $X$ to its mirror Calabi-Yau $\check{X}$, type IIA theory to type IIB, topological A-model to B-model, and D6 branes on a special Lagrangian three-cycle $L\subset X$ to D5 branes on a holomorphic submanifold (typically curve) $\mathcal{C}\subset \check{X}$, which is a mirror partner of $L$. 
Under the mirror map two D-brane superpotentials for the low-energy effective theory on the branes are related to each other. 
Thus mirror symmetry insists that the disk invariants in the A-model can be computed by the B-model calculation, which is rather easy because there is no instanton contributions. 
More concretely, the open mirror conjecture (in the on-shell $\mathbb{Z}_2$-vacua case) insists that 
the B-model domainwall tension $\mathcal{T}_B(z)$, 
which depends only on a complex structure moduli $z$ of $\check{X}$ (now open moduli of B-brane is discrete and does not appear explicitly), can be interpreted as the quantity in the A-model side $\mathcal{T}_A(t)$, 
where $t$ is a complexified K\"ahler moduli of $X$: 
\beq
\frac{\mathcal{T}_B(z)}{\varpi_0(z)}=\mathcal{T}_A(t)
=\sum_{\text{d:odd}} N_d^0q^{d/2}=\sum_{\text{d:odd,\\k:odd}}\frac{2n_d^{(0,\text{real})}}{k^2}q^{dk/2} \ \ \ (q=e^{2\pi it}). \label{A-T}
\eeq
Here $t=\frac{\varpi_1(z)}{\varpi_0(z)}$ is the mirror map, 
$\varpi_0(z)$ is the fundamental period and $\varpi_1(z)$ is the logarithmic period. 
$N_d^{0}$ are open Gromov-Witten invariants and $n_d^{(0,\text{real})}$ are real BPS invariants (the number of real rational curves, and 
half of the number of holomorphic disks \cite{PaSoWa}) which can be obtained after re-summation by Ooguri-Vafa multi-covering formula \cite{OV}. 
We expect that the latter invariants must be integers by the intuitive geometric pictures, namely, 
they are the number of certain geometric objects in Calabi-Yau.

Usually we take B-brane (holomorphic curves) as an intersection locus of two hyperplanes $Q_1$ and $Q_2$ with Calabi-Yau $\check{X}=\cap_i\{P_i=0\}$, 
\beq
Q_1\cap Q_2 \cap \check{X}. 
\eeq
For all of the previously studied cases, the B-branes have the above structure in common, at least in the on-shell situation \cite{MW,KW1,KS1,Walcher}. 
The obtained holomorphic curves consist of typically two irreducible components and 
they correspond to two vacua\footnote{
Some Calabi-Yau manifolds, typically ones defined in a weighted projective space, 
have branes with $\mathbb{Z}_k$-vacua $(k>2)$ \cite{Walcher}. 
It is possible that some of pfaffians we treat in this paper have this type of vacua, but we do not consider such possibilities in this paper. 
}. 
We assume that there exists a mirror partner A-brane in mirror Calabi-Yau $X$ and it has discrete $\mathbb{Z}_2$ brane moduli. 
We also assume that by open mirror symmetry the enumerative prediction of disk invariants ending such A-brane geometry can be obtained. 
Although this procedure and assumption are somewhat ad-hoc, we can certainly obtain integral invariants by the enumerative predictions, as shown in later sections.

Later, we will compute the superpotentials of such B-branes in the pfaffian Calabi-Yau manifolds, by applying our method proposed recently \cite{FNSS}.

\subsection{Pfaffian Calabi-Yau manifold}
In this subsection we review some basic facts about pfaffian Calabi-Yau variety. 
Some models which will be treated in later sections are introduced. 
From now on, we mainly use the notation in \cite{Kanazawa}. 

\subsubsection{Pfaffian Calabi-Yau $3$-fold}
The pfaffian is defined for a skew-symmetric matrix. 
Let $N=\{n_{i,j}\}$ be a $2r\times 2r$ skew-symmetric matrix. 
The pfaffian of $N$ is defined by the following equation: 
\beq
\mathrm{Pf}(N) = \frac{1}{2^r r!}\sum_{\sigma\in S_{2r}}\mathrm{sgn}(\sigma)\prod_{i=1}^{r}n_{\sigma(2i-1),\sigma(2i)}, 
\eeq
where $S_{2r}$ is the symmetric group and $\mathrm{sgn}(\sigma)$ is the signature of $\sigma$. 
For the $(2r+1)\times (2r+1)$ matrix, the pfaffian is defined to be zero. 
The pfaffian is equal to the square root of the determinant. 
We also define $P_{i_1i_2...i_l}:=\text{Pf}(N_{i_1i_2...i_l})$ where $N_{i_1i_2...i_l}$ is a skew symmetric matrix obtained by removing all the $i_j$-th rows and columns from $N$. 

For the definition of pfaffian variety, we need the date $(\mathcal{E},N)$ defined below. 
In this paper, we consider the case defined in a single weighted projective space $\mathbb{P}_{w}^n$ $(n>3)$. 
Let $\mathcal{E}$ be a locally free sheaf of odd rank $2r+1$ on $\mathbb{P}^n_{w}$, 
and $t\in\mathbb{Z}$. 
Then, a generically-taken global section $N\in H^0(\mathbb{P}^n_{w},\wedge^2\mathcal{E}(t))$ defines an alternating morphism 
\beq
\mathcal{E}^{\vee}(-t)\stackrel{N}{\longrightarrow}\mathcal{E}. 
\eeq
Then we have a following exact complex, called a pfaffian complex associated to the data $(\mathcal{E},N)$: 
\beq
0\rightarrow
\mathcal{O}_{{\mathbb{P}^n_{w}}}(-t-2s)
\stackrel{{}^tP}{\longrightarrow}
\mathcal{E}^{\vee}(-t-s)
\stackrel{N}{\longrightarrow}
\mathcal{E}(-s)
\stackrel{P}{\longrightarrow}
\mathcal{O}_{{\mathbb{P}^n_{w}}}, 
\eeq
where $s=\int_{\text{P.D.}(H)}c_1(\mathcal{E})+rt$ ($H$: hyperplane section) and $P$ is 
\beq
P:=\frac{1}{r!}\wedge^r N\in H^0(\mathbb{P}^n_{w},\wedge^{2r}\mathcal{E}(rt)). 
\eeq
The first and the third morphisms are given by taking the wedge product with $P$ and ${}^tP$ respectively. 
Note that, once we fix a framing ${\{e_i\}}_{i=1}^{2r+1}$ of $\mathcal{E}$, 
$N$ is just a matrix and 
\beq
P=\sum_{i=1}^{2r+1}P_i\wedge_{j\neq i}e_j
=\sum_{i=1}^{2r+1}\text{Pf}(N_i)\wedge_{j\neq i}e_j. 
\eeq
Then, the projective variety $X\subset \mathbb{P}^n$ is called a pfaffian variety associated to the data $(\mathcal{E},N)$ if the structure sheaf $\mathcal{O}_X$ is given by $Coker(N)$. 
The sheaf $Im(P)\subset\mathcal{O}_{\mathbb{P}^n}$ is called the pfaffian ideal sheaf and denoted by $\mathcal{I}_X$. 
The pfaffian variety is not complete intersection globally but locally so. 
On the affine open local patch where $P_{\nu_0 \nu_1 \nu_2}\neq 0$, 
this variety can be expressed as the complete intersection of $\{P_{\nu_i}\}_{i=0,1,2}$.

Let us consider the special case, 
$3$-dimensional Calabi-Yau variety $X_i$ in a $6$-dimensional weighted projective space, $\mathbb{P}^6_{w}$ 
(Here, $w=[w_i]_{i=0,...,6}$ denotes weights). 
In this case, $t+2s=|w|:=\sum_{i}w_i$, 
and we may assume $t=1$ for $|w|$ odd and $t=0$ for $|w|$ even. 
The pfaffian complex can be written as follows: 
\beq
0\rightarrow
\mathcal{O}_{\mathbb{P}^6_{w}}(-7)
\stackrel{{}^tP}{\longrightarrow}
\mathcal{E}^{\vee}(-4)
\stackrel{N}{\longrightarrow}
\mathcal{E}(-3)
\stackrel{P}{\longrightarrow}
\mathcal{O}_{\mathbb{P}^6_{w}}
\rightarrow
\mathcal{O}_{X}
\rightarrow
0. 
\eeq
$X_i$ is Calabi-Yau since the canonical sheaf of this variety is trivial 
\beq
\omega_{X_i}\cong\mathcal{E}xt^3(\mathcal{O}_{X_i},\omega_{\mathbb{P}_{w}^6})\cong \mathcal{O}_{X_i}. 
\eeq

Some pfaffian Calabi-Yau manifolds in $\mathbb{P}^6$, whose degrees are from $11$ to $17$, are constructed by F. Tonoli in \cite{Tonoli}. 
It is showed that the degree $11$ case is not smooth and 
the degree $12$ case can be expressed as the complete intersections of two degree $2$ equations and one degree $3$ equation in $\mathbb{P}^{6}$. 
So from non-complete intersection viewpoint, the next interesting one is the degree $13$ case. 
This Calabi-Yau is one-moduli, i.e. $h^{1,1}=1$. 
The closed mirror symmetry for this Calabi-Yau was discussed in detail in \cite{Kanazawa}. 
By extension, new Pfaffian Calabi-Yau $3$-folds with $h^{1,1}=1$ in a weighted projective space (and their expected mirror partners) were constructed in \cite{Kanazawa}. 
In the next subsection, we will discuss some of them.

\subsubsection{Models and Mirror families}

In this paper, we treat $X_i$ ($i=13$, $5$, $7$, $10$). 
The subscript denotes its degree, namely, 
\beq
i=\int_{X_i}H^3, 
\eeq
where $H$ is the hyperplane section. 
All of them are defined in a $6$-dimensional weighted projective space, 
$X_{i}\subset \mathbb{P}_{w}^6$. 
They are expressed as the degeneracy locus of a generic alternating morphism $\mathcal{E}^{\vee}(-1)\stackrel{N}{\longrightarrow}\mathcal{E}$ and 
defined commonly by the $5\times 5$ pfaffians. 
The ambient projective space and the bundle where section lives are as follows: 
\begin{align}
&\text{For} \ i=13, \ \ \ \ w=(1^7) \ \ \text{and} \ \ 
\mathcal{E}=\mathcal{O}_{\mathbb{P}^6}(1)\oplus\mathcal{O}_{\mathbb{P}^6}^{\oplus 4}. \\
&\text{For} \ i=5, \ \ \ \ w=(1^4,2^3) \ \ \text{and} \ \ 
\mathcal{E}=\mathcal{O}_{\mathbb{P}_{w}}^{\oplus 5}(1). \\
&\text{For} \ i=7, \ \ \ \ w=(1^5,2^2) \ \ \text{and} \ \ 
\mathcal{E}=\mathcal{O}_{\mathbb{P}_{w}}^{\oplus 2}(1)\oplus\mathcal{O}_{\mathbb{P}_{w}}^{\oplus 3}. \\
&\text{For} \ i=10, \ \ \ \ w=(1^6,2^1) \ \ \text{and} \ \ 
\mathcal{E}=\mathcal{O}_{\mathbb{P}_{w}}^{\oplus 4}(1)\oplus\mathcal{O}_{\mathbb{P}_{w}}. 
\end{align}
In \cite{Rodland,Kanazawa} it was verified that for a generic choice of $N$, all of the above pfaffians are $3$-dimensional {\it smooth} Calabi-Yau varieties. 

Topologically, Calabi-Yau manifold is specified by two data, $h^{1,1}$ and $h^{2,1}$, 
here $h^{p,q}$ is the Hodge number (the rank of $H^{p,q}$). 
We are mainly interested in the one with the lowest number of K\"ahler moduli, $h^{1,1}=1$. 
All of $X_i$ we treat in this paper have this property. 
We list some geometric data of them in Table \ref{geometric data}. 
These quantities were computed in \cite{Rodland} for $X_{13}$ and in \cite{Kanazawa} for $X_5$, $X_7$ and $X_{10}$. 
\begin{table}
\begin{center}
\begin{tabular}{c|ccccc} \toprule
$X_i$ & degree ($H^3$) & $h^{1,1}$ & $h^{2,1}$ & $\chi$ & $c_2\bullet H$ \\ 
\midrule
$X_{13}$ & 13 & 1 & 61 & -120 & 58 \\ [-5pt] 
$X_5$ & 5 & 1 & 51 & -100 & 38 \\ [-5pt] 
$X_7$ & 7 & 1 & 61 & -120 & 46 \\ [-5pt] 
$X_{10}$ & 10 & 1 & 59 & -116 & 52 \\ 
\bottomrule
\end{tabular}
\end{center}
\caption{Geometric data of pfaffians}\label{geometric data}
\end{table}
The existence of smooth Calabi-Yau 3-folds with the above topological invariants was previously conjectured in \cite{data} from the viewpoint of Calabi-Yau equations. 

Then we turn to consider the mirror partners of them. 
We denote the mirror partner of $X_{i}$ as ${(\check{X}_i)}_{\psi}$ (or simply, $\check{X}_i$), 
where $\psi\in \mathbb{P}^1$ is a parameter of complex structure moduli. 
The relation to the parameter $t$ used in \cite{Kanazawa} is $\psi=1/t$ and the words {\it large moduli} and {\it small moduli} are used with respect to this parameter $\psi$. 
Because they are not complete intersections, 
we cannot apply the Batyrev-Borisov mirror construction to these models. 
Alternatively the mirror construction via the tropical geometry was proposed and the candidate of mirror partner of $X_{13}$ was constructed in \cite{Boehm} and the candidates of mirror partners of $X_5$, $X_7$ and $X_{10}$ were constructed in \cite{Kanazawa}\footnote{
They are constructed by a certain orbifolding, 
and generic members of these one-parameter families are quite singular. 
We expect that crepant resolutions of them are correct mirror families. 
However such resolutions are not found yet. 
It is expected that we can avoid this problem by our prescription of analytic continuations explained in section \ref{formal}. 
}. 
The confirmation that they are mirror partners were examined by comparison of some geometric data. 
The corresponding Picard-Fuchs equations were computed and we can find the expected equations listed in \cite{data}. 
The mirror Calabi-Yaus $\check{X}_i$ certainly have {\it mirrored} Hodge numbers and Euler characteristic, namely, 
$h^{2,1}(\check{X}_i)=h^{1,1}(X_i)=1$, 
$h^{1,1}(\check{X}_i)=h^{2,1}(X_i)$, 
and $\chi(\check{X}_i)=-\chi(X_i)$. 
In the following, we list explicit forms of skew-symmetric matrices given in \cite{Kanazawa,Boehm} and pfaffian ideal sheaves of them. 
They have commonly good structures, 
in the sense that the hyperplane equations appear explicitly in the defining equations of Calabi-Yau manifolds.

For $(\check{X}_{13})_{\psi}$, 
the ambient space is $\mathbb{P}^6/\mathbb{Z}_{13}$ 
and $\mathbb{Z}_{13}$ acts as follows: 
\beq
\zeta_{13}\bullet[x_0:x_1:x_2:x_3:x_4:x_5:x_6]=[x_0:\zeta_{13}^4 x_1:\zeta_{13}^8 x_2:\zeta_{13}^{10}x_3:\zeta_{13}^{10}x_4:\zeta_{13}^{11}x_5:\zeta_{13}^{11}x_6], 
\eeq
where the generator of $\mathbb{Z}_{13}$ is given by $\zeta_{13}=\exp\left(\frac{2\pi i}{13}\right)$. 
The skew symmetric matrix is 
\begin{align}
(N_{13})_{\psi}=
\begin{pmatrix}
0 & \frac{1}{\psi}x_0^2 & x_5x_6 & x_3x_4 & \frac{1}{\psi}x_2^2 \\
-\frac{1}{\psi}x_0^2 & 0 & \frac{1}{\psi}(x_3+x_4) & x_2 & x_1 \\
-x_5x_6 & -\frac{1}{\psi}(x_3+x_4) & 0 & \frac{1}{\psi}x_1 & x_0 \\
-x_3x_4 & -x_2 & -\frac{1}{\psi}x_1 & 0 & \frac{1}{\psi}(x_5+x_6) \\
-\frac{1}{\psi}x_2^2 & -x_1 & -x_0 & -\frac{1}{\psi}(x_5+x_6) & 0 
\end{pmatrix}. \label{m13}
\end{align}
The pfaffian ideal sheaf of this family,  $\mathcal{I}_{\check{X}_{13}}=\langle P_0,P_1,P_2,P_3,P_4\rangle$, is generated by the following polynomials: 
\begin{align}
&P_0=x_0x_2-\frac{1}{\psi}x_1^2-\frac{1}{\psi^2}(x_3+x_4)(x_5+x_6),\\
&P_1=x_0x_3x_4-\frac{1}{\psi}x_5x_6(x_5+x_6)-\frac{1}{\psi^2}x_1x_2^2,\\
&P_2=x_1x_3x_4-\frac{1}{\psi}x_2^3-\frac{1}{\psi^2}x_0^2(x_5+x_6),\\
&P_3=x_1x_5x_6-\frac{1}{\psi}x_0^3-\frac{1}{\psi^2}x_2^2(x_3+x_4),\\
&P_4=x_2x_5x_6-\frac{1}{\psi}x_3x_4(x_3+x_4)-\frac{1}{\psi^2}x_0^2x_1. 
\end{align}
In \cite{Kanazawa} it was observed that $z=\psi^{-7}\in\mathbb{P}^1$ is the genuine moduli parameter. 

For $(\check{X_5})_{\psi}$, 
the ambient space is $\mathbb{P}^6_{(1^4,2^3)}/\mathbb{Z}_{10}$ and 
$x_0$, $x_1$, $x_2$ and $x_3$ are weight $1$, and $x_4$, $x_5$ and $x_6$ are weight $2$. 
The orbifold group acts as 
\beq
\zeta_{10}\bullet[x_0:x_1:x_2:x_3:x_4:x_5:x_6]=[x_0:x_1:\zeta_{10}x_2:\zeta_{10}x_3:\zeta_{10}^{4}x_4:\zeta_{10}^{6}x_5:\zeta_{10}^{3}x_6], 
\eeq
where $\zeta_{10}$ is the generator of $\mathbb{Z}_{10}$ and given by $\zeta_{10}=\exp\left(\frac{2\pi i}{10}\right)$. 
The skew symmetric matrix is 
\begin{align}
(N_5)_{\psi}=
\begin{pmatrix}
0 & \frac{1}{\psi}x_6 & x_4 & x_0x_1 & \frac{1}{\psi}x_5 \\
-\frac{1}{\psi}x_6 & 0 & \frac{1}{\psi}(x_0^2+x_1^2) & x_5 & x_2x_3 \\
-x_4 & -\frac{1}{\psi}(x_0^2+x_1^2) & 0 & \frac{1}{\psi}(x_2^2+x_3^2) & x_6 \\
-x_0x_1 & -x_5 & -\frac{1}{\psi}(x_2^2+x_3^2) & 0 & \frac{1}{\psi}x_4 \\
-\frac{1}{\psi}x_5 & -x_2x_3 & -x_6 & -\frac{1}{\psi}x_4 & 0 
\end{pmatrix}. \label{m5}
\end{align}
The pfaffian ideal sheaf of this family is generated by the following polynomials: 
\begin{align}
&P_0=-x_5x_6+\frac{1}{\psi}x_2x_3(x_2^2+x_3^2)+\frac{1}{\psi^2}(x_0^2+x_1^2)x_4, \\
&P_1=-x_0x_1x_6+\frac{1}{\psi}x_4^2+\frac{1}{\psi^2}(x_2^2+x_3^2)x_5, \\
&P_2=-x_0x_1x_2x_3+\frac{1}{\psi}x_5^2+\frac{1}{\psi^2}x_4x_6, \\
&P_3=-x_2x_3x_4+\frac{1}{\psi}x_6^2+\frac{1}{\psi^2}(x_0^2+x_1^2)x_5, \\
&P_4=-x_4x_5+\frac{1}{\psi}x_0x_1(x_0^2+x_1^2)+\frac{1}{\psi^2}(x_2^2+x_3^2)x_6. 
\end{align}
In \cite{Kanazawa} it was observed that $z=\psi^{-10}\in\mathbb{P}^1$ is the genuine moduli parameter. 
It is worth noting that, in this case, the forms of the {\it hyperplanes} in the defining equations are different from the other cases, i.e., the degree two form\footnote{
Thus the use of terminology {\it hyperplane} is not suitable and we use italic letters for this reason. 
Precisely the hyperplanes in this case are expressed as $\{x_0+\alpha x_1=x_2+\beta x_3=0\}$ $(\alpha^2=\beta^2=-1)$ and this is important in the later arguments in section \ref{5}. 
}. 

For $(\check{X_7})_{\psi}$, 
the ambient space is $\mathbb{P}_{(1^5,2^2)}/\mathbb{Z}_{7}$, where $x_0$, $x_1$, $x_2$, $x_3$ and $x_4$ are weight $1$, and $x_5$ and $x_6$ are weight $2$. 
The orbifold group acts as 
\beq
\zeta_{7}\bullet[x_0:x_1:x_2:x_3:x_4:x_5:x_6]=[x_0:x_1:\zeta_{7}^4 x_2:\zeta_{7}x_3:\zeta_{7}x_4:\zeta_{7}^{3}x_5:\zeta_{7}^{6}x_6], 
\eeq
where the generator is given by $\zeta_{7}=\exp\left(\frac{2\pi i}{7}\right)$. 
The skew symmetric matrix is 
\begin{align}
(N_7)_{\psi}=
\begin{pmatrix}
0 & \frac{1}{\psi}x_2^3 & x_0x_1 & x_5 & \frac{1}{\psi}x_6 \\
-\frac{1}{\psi}x_2^3 & 0 & \frac{1}{\psi}x_5 & x_6 & x_3x_4 \\
-x_0x_1 & -\frac{1}{\psi}x_5 & 0 & \frac{1}{\psi}(x_3+x_4) & x_2 \\
-x_5 & -x_6 & -\frac{1}{\psi}(x_3+x_4) & 0 & \frac{1}{\psi}(x_0+x_1) \\
-\frac{1}{\psi}x_6 & -x_3x_4 & -x_2 & -\frac{1}{\psi}(x_0+x_1) & 0 
\end{pmatrix}. \label{m7}
\end{align}
The pfaffian ideal sheaf of this family is generated by the following polynomials: 
\begin{align}
&P_0=-x_2x_6+\frac{1}{\psi}x_3x_4(x_3+x_4)+\frac{1}{\psi^2}(x_0+x_1)x_5, \\
&P_1=-x_2x_5+\frac{1}{\psi}x_0x_1(x_0+x_1)+\frac{1}{\psi^2}(x_3+x_4)x_6, \\
&P_2=-x_3x_4x_5+\frac{1}{\psi}x_6^2+\frac{1}{\psi^2}(x_0+x_1)x_2^3, \\
&P_3=-x_0x_1x_3x_4+\frac{1}{\psi}x_2^4+\frac{1}{\psi^2}x_5x_6, \\
&P_4=-x_0x_1x_6+\frac{1}{\psi}x_5^2+\frac{1}{\psi^2}x_2^3(x_3+x_4). 
\end{align}
In \cite{Kanazawa} it was observed that $z=\psi^{-9}\in\mathbb{P}^1$ is the genuine moduli parameter. 

For $(\check{X_{10}})_{\psi}$, 
the ambient space is $\mathbb{P}_{(1^6,2^1)}/\mathbb{Z}_{10}$, and 
$x_0$, $x_1$, $x_2$, $x_3$, $x_4$ and $x_5$ have weight $1$ and $x_6$ has weight $2$. 
The orbifold group act as 
\beq
\zeta_{10}\bullet[x_0:x_1:x_2:x_3:x_4:x_5:x_6]=[x_0:x_1:\zeta_{10}^{6} x_2:\zeta_{10}^{6}x_3:\zeta_{10}^{9}x_4:\zeta_{10}^{7}x_5:\zeta_{10}^{1}x_6], 
\eeq
where the generator of $\mathbb{Z}_{10}$ is given by $\zeta_{10}=\exp\left(\frac{2\pi i}{10}\right)$. 
The skew symmetric matrix is 
\begin{align}
(N_{10})_{\psi}=
\begin{pmatrix}
0 & \frac{1}{\psi}x_2^4 & x_0x_1 & x_6 & \frac{1}{\psi}(x_2+x_3) \\
-\frac{1}{\psi}x_2^4 & 0 & \frac{1}{\psi}x_6 & x_2x_3 & x_5 \\
-x_0x_1 & -\frac{1}{\psi}x_6 & 0 & \frac{1}{\psi}x_5^2 & x_4 \\
-x_6 & -x_2x_3 & -\frac{1}{\psi}x_5^2 & 0 & \frac{1}{\psi}(x_0+x_1) \\
-\frac{1}{\psi}(x_2+x_3) & -x_5 & -x_4 & -\frac{1}{\psi}(x_0+x_1) & 0 
\end{pmatrix}. \label{m10}
\end{align}
The pfaffian ideal sheaf of this family is generated by the following polynomials: 
\begin{align}
&P_0=-x_2x_3x_4+\frac{1}{\psi}x_5^3+\frac{1}{\psi^2}(x_0+x_1)x_6, \\
&P_1=-x_4x_6+\frac{1}{\psi}x_0x_1(x_0+x_1)+\frac{1}{\psi^2}(x_2+x_3)x_5^2, \\
&P_2=-x_5x_6+\frac{1}{\psi}x_2x_3(x_2+x_3)+\frac{1}{\psi^2}(x_0+x_1)x_4^2, \\
&P_3=-x_0x_1x_5+\frac{1}{\psi}x_4^3+\frac{1}{\psi^2}(x_2+x_3)x_6, \\
&P_4=-x_0x_1x_2x_3+\frac{1}{\psi}x_6^2+\frac{1}{\psi^2}x_4^2x_5^2. 
\end{align}
In \cite{Kanazawa} it was observed that $z=\psi^{-8}\in\mathbb{P}^1$ is the genuine moduli parameter. 

In \cite{Kanazawa} it was observed that the orbifold groups of the above models are $\mathbb{Z}_{|w|}$ and the genuine moduli parameters are $\psi^{-|w|}$, where $|w|=\sum_{i=0}^6w_i$. 
We do not know the clear reason behind this fact.

\subsubsection{Holomorphic $3$-form and Period}\label{hol3_period}

Every smooth Calabi-Yau $3$-fold has the holomorphic $3$-form which vanishes nowhere, 
as a global section of $\omega_{\check{X}} \cong \Omega_{\check{X}}^3$. 
This $3$-form is unique up to multiplication with a non-zero constant. 
The pfaffian variety cannot be expressed as a complete intersection and there is no way of explicitly getting one in general. 
However we have an analogous way of obtaining it in pfaffian cases, as R\o dland shown in \cite{Rodland}. 

Let $\nu$ be a permutation of the elements of $\mathbb{Z}_{5}$. 
Then the holomorphic $3$-form on $\check{X}\vert_{\{P_{\nu_0\nu_1\nu_2}\neq 0\}}\subset\mathbb{P}_{w}$ can be expressed as 
\beq
\Omega^{3,0}=\frac{\vert G\vert}{(2\pi i)^6}\text{Res}_{\check{X}}\frac{P_{\nu_{0}\nu_{1}\nu_{2}}}{P_{\nu_{0}}P_{\nu_{1}}P_{\nu_{2}}}\omega_{0}=\frac{\vert G\vert}{(2\pi i)^6}\int_{T_\epsilon(\check{X})}\frac{P_{\nu_0\nu_1\nu_2}}{P_{\nu_0}P_{\nu_1}P_{\nu_2}}\omega_0, \label{hol3form}
\eeq
where 
\beq
\omega_0=\sum_{i=0}^6(-1)^{i}w_ix_idx_0\wedge...\wedge \Hat{dx_i} \wedge ...\wedge dx_6. 
\eeq
Here, $\int_{T_\epsilon(\check{X})}$ means the integral over a small tube encircling $3$-fold locus $\{P_{\nu_0}=0\}\cap\{P_{\nu_1}=0\}\cap\{P_{\nu_2}=0\}$ 
and $\vert G \vert$ is the order of the orbifold group.

The periods are defined as the integrals of this $3$-form over $3$-dimensional subspaces in Calabi-Yau. 
For the fundamental period $\Pi_0$ (resp. the domainwall tension $\mathcal{T}$), $3$-dimensional subspace is 
a suitably chosen $3$-cycle (resp. a $3$-chain whose boundary is a B-brane locus), namely, 
\beq
\Pi_0 \ (\text{or }\mathcal{T}) \ =\int_{\text{\{$3$-cycle or chain\}}}\Omega^{3,0}=
\frac{|G|}{(2\pi i)^6}\int_{T_{\epsilon}(\check{X}\cap\text{\{$3$-cycle or chain\}})}\frac{P_{\nu_0\nu_1\nu_2}}{P_{\nu_0}P_{\nu_1}P_{\nu_2}}\omega_0. \label{period} 
\eeq
In later sections, 
we evaluate the above period integrals directly. 
In cases treated in this paper, 
the ambient space is a complex $6$-dimensional manifold and 
the above integral contains three residue integrals and the integral over real $3$-dimensional cycle/chain. 
Thus the integrals \eqref{period} are expressed as $6$ residue integrals, 
or $5$ residue integrals and a $1$ line integral, 
in a $6$-dimensional ambient (weighted) projective space, 
as explained in later sections.

In $\mathbb{Z}_2$-vacua case, 
the tension of BPS domainwall defined by the above integral has the form obtained by shifting $n\longrightarrow n+1/2$ in the fundamental period (up to overall normalization): 
\beq
\Pi_0=\sum_{n=0}^{\infty} a(n)z^n\longrightarrow \mathcal{T}\sim\sum_{n=0}^{\infty} a(n+1/2)z^{n+1/2}, 
\eeq
where $z$ is a complex structure moduli, 
$a(n)$ are certain coefficients, 
and $n$ runs over all non-negative integers. 
See e.g. \cite{Wa1,MW,KW1,KS1}. 
This is related to the fact that the domainwall tension is one of the solutions of the following modified Picard-Fuchs equation \cite{Wa1}: 
\beq
(2\theta+1)\mathcal{L}_{\text{PF}}\mathcal{T}=0, 
\eeq 
where $\mathcal{L}_{\text{PF}}$ is the ordinary Picard-Fuchs operator. 
Namely, this is one of the solutions of the inhomogeneous Picard-Fuchs equation, whose inhomogeneous term is in the following form: 
\beq
\mathcal{L}_{\text{PF}}\mathcal{T}=\text{Const.}z^{1/2}\label{inhomogeneous1}. 
\eeq
We should be careful with the $1/2$-shift in the case of the pfaffian Calabi-Yaus treated in the present paper. 
In this case, the several periods have one more summation (with respect to $k$), in addition to the ordinary summation (with respect to $n$), whose range runs from $0$ to a finite integer if we fix $n$. 
In particular, the fundamental period takes the following form: 
\beq
\Pi_0=\sum_{n=0}^{\infty}\left(a(n)\sum_{k=0}^{f(n)}b(n,k)\right)z^n. \label{ksum}
\eeq
Moreover, the fundamental period has several different-looking forms that are related 
by the hypergeometric identities \cite{binomial}. 
It sometimes happens that, although two such formulas are identical in the sense of the fundamental period, 
domainwall tensions obtained by the above simple shift do not coincide. 
The reason for this is that the hypergeometric identities may hold only in the case where the arguments of the Gamma functions are all integers. 
Thus, if we choose an inappropriate form, the summation of $k$ does not stop at a finite value after the $1/2$-shift. 
This is not desirable. We must be careful about this. 

In our method, which will be discussed in section \ref{formal}, 
the different representations of the same period resulting from the hypergeometric identities can be explained simply as 
how differently the local coordinates of the period integrals have been assigned. 
See \eqref{assign} and the explanation below. 
The additional summation with respect to $k$ can be also understood intuitively as the existence of the binomial expansion in the integral formula in our formalism. 
See e.g., \eqref{13binomial2} and \eqref{13binomial1}. 

Furthermore, interestingly, we observe that the above simple inhomogeneous term \eqref{inhomogeneous1} changes in the case of the pfaffians we treat in this paper. 
The modification turns out to be the following form: 
\beq
\mathcal{L}_{\text{PF}}\mathcal{T}=c_1z^{1/2}+c_2z^{3/2}+c_3z^{5/2}, \label{inh}
\eeq
where $c_i$ $(i=1,2,3)$ are constants. 
Thus the computation of the inhomogeneous term may be very complicated and our direct computation method of the period integrals is advantageous in the present pfaffian cases. 
The above form of the inhomogeneous term can be understood naturally since the pfaffians we treat have the following Picard-Fuchs operators: 
\begin{align}
\mathcal{L}_{\text{PF}}=&A\Theta^4+Bz(b_1\Theta^4+b_2\Theta^3+...)\nonum\\
&+Cz^2(c_1\Theta^4+c_2\Theta^3+...)
+Dz^3(d_1\Theta^4+d_2\Theta^3+...)
+Ez^4(e_1\Theta^4+e_2\Theta^3+...), 
\end{align}
where $\Theta=z\partial_z$ is the Euler operator, and $A$, $B$, ..., are certain numerical constants. 
For the concrete expressions of the present models, see \cite{data,Kanazawa}.  When this operator acts on the chain integral, we have the boundary contributions with different powers of $z$ which result in the inhomogeneous terms \eqref{inh}. 
In Table \ref{inhomogeneous} in section \ref{inhomo}, 
we list the inhomogeneous terms of the pfaffian Calabi-Yaus we treat.

\subsubsection{Comment on Moduli space}
It is interesting that some of the pfaffian Calabi-Yaus have two maximal unipotent monodromy points in its complex structure moduli space, where the geometry is maximally degenerating or almost so. 
Each such point corresponds to the large volume limit points of different Calabi-Yaus in the A-model interpretation under the mirror map. 
Thus surprisingly the complexified K\"{a}hler moduli spaces of two different A-side Calabi-Yaus $X$ and $X'$ are governed by the same {\it single} complex moduli space in the mirror B-side, $\check{X}_{\psi}$, 
and described as two different limits of a {\it single} Picard-Fuchs operator (conjecture noted in \cite{Rodland}). 
In this paper, we call these two limits as {\it large moduli limit} and {\it small moduli limit}, because in the B-model side they are realized as two limits of $\psi$, namely, $\psi\longrightarrow\infty$ and $\psi\longrightarrow 0$. 
Using the mirror map and going to the A-model interpretation, 
we can make the instanton expansions at each degeneration point. 
Our main objective in this paper is to carry out the instanton calculations of open string sector, i.e., the disk instanton counting, by using open mirror symmetry. 

The models we treat have the following differences. 
\begin{itemize}
\item The mirror of the degree $13$, $(\check{X}_{13})_{\psi}$, has two special points in its moduli space. 
The large moduli limit is the model No.99 listed in \cite{data}. 
The predicted model of the small moduli limit is the model No.225 listed in \cite{data}, 
but we do not know the concrete construction of the A-model geometry. 

\item The mirror of the degree $5$, $(\check{X}_5)_{\psi}$, is the model No.302 listed in \cite{data}. 
This also has two special points in its moduli space, but both correspond to the same Calabi-Yau. 
So we call this model {\it self-dual}. 

\item The mirror of the degree $7$, $(\check{X}_7)_{\psi}$, 
which is No.109 model listed in \cite{data}, 
has only one special point in its moduli space. 
The small moduli limit point is not a maximally unipotent point. 

\item The mirror of the degree $10$, $(\check{X}_{10})_{\psi}$, has two special points in its moduli space. 
The large moduli limit is the model No.263 listed in \cite{data}. 
The small moduli limit is the model No.271 listed in \cite{data}, 
but we do not know the concrete construction. 
\end{itemize}

Some quantities of the pfaffians are listed in Table \ref{moduli}. 
\begin{table}
\begin{center}
\begin{tabular}{l|ccccc} 
\toprule
$\check{X}_i$ & limit & No. in \cite{data} & (diss$1$) & (diss$2$) & genuine moduli \\ 
\midrule
$\check{X}_{13}$ & large & 99 & $1-349z-256z^2$ & $13-16z$ & $z=\psi^{-7}$ \\ [-5pt] 
 & small & 225 & $1-89344x-16777216x^2$ & $1+53248x$ & $x=-1/(2^{16}\psi^{-7})$ \\ [-5pt] 
\midrule
$\check{X}_5$ & large & 302 & $1-1968z+256z^2$ & $5(1-16z)$ & $z=\psi^{-10}$ \\ [-5pt] 
 & small & 302 (self-dual) & $1-1968x+256x^2$ & $5(1-16x)$ & $x=-1/(2^8\psi^{-10})$ \\ [-5pt] 
\midrule
$\check{X}_7$ & large & 109 & $1-1080z-432z^2$ & $7-36z$ & $z=\psi^{-9}$ \\ [-5pt] 
\midrule
$\check{X}_{10}$ & large & 263 & $1-544z+256z^2$ & $2(5-16z)$ & $z=\psi^{-8}$ \\ 
 & small & 271 & $1-8704x+65536x^2$ & $2(1-1280x)$ & $x=-1/(2^{12}\psi^{-8})$ \\ 
\bottomrule
\end{tabular}
\end{center}
\caption{Several data at two moduli limits}\label{moduli}
\end{table}
Here, $(\text{diss}1)$ is the discriminants of pfaffian Calabi-Yaus, and 
Calabi-Yau is degenerate and has double points at its roots. 
$(\text{diss}2)$ is the numerator of the quantum Yukawa coupling (see \eqref{quantum_yukawa}) and 
its roots also correspond to singular points of the Picard-Fuchs operator. 

In \cite{Kanazawa}, the analyses of (closed) mirror symmetry for the present pfaffian Calabi-Yaus and the predictions of the number of rational curves were done. 
The study in the small moduli limit was done by simply transforming the Picard-Fuchs equations from the large moduli limit to the small moduli limit and 
it was observed that a certain combination of moduli is suitable from the viewpoint of the enumerative predictions. 
In Table \ref{moduli}, we show such combinations in terms of $\psi$ used in \eqref{m13}, \eqref{m5}, \eqref{m7} and \eqref{m10}. 
It was also observed that the shift $\Theta\longrightarrow \Theta+1/2$ is needed 
to obtain the correct Picard-Fuchs operators in the small moduli limit. 
In our method, 
it is possible to understand why such moduli combination and the shift $\Theta\longrightarrow \Theta+1/2$ are needed. 
\\

Finally, before closing this section, we comment briefly the degree $14$ case, which is the most studied example from closed mirror symmetry viewpoint \cite{Rodland,HoKo}. 
It was also studied from the viewpoint of the GLSM with non-abelian gauge group \cite{HT}. 
The direct computation of Gromov-Witten invariants was done in \cite{Tjotta}. 
Although this model has a very interesting and beautiful structure, 
we do not treat this model in this paper. 
This model is also $1$-moduli Calabi-Yau and is interesting in the sense that the A-model side geometries of both of the two moduli limits are explicitly known, 
while the A-model geometries of the small moduli limit of the models treated in the present paper are not known. 

The A-model side of the large moduli limit of the degree $14$ case is expressed by a $7\times 7$ pfaffian, 
\beq
X_{1}=\text{Pf}(7)\cap\check{\mathbb{P}}^6\subset\check{\mathbb{P}}^{20} 
\eeq
and that of the small moduli limit is expressed by a certain Grassmannian, 
\beq
X_{2}=\text{Gr}(2,7)\cap\mathbb{P}^{13}\subset\mathbb{P}^{20}. 
\eeq
Here, $\check{\mathbb{P}}^{20}$ is the dual projective to $\mathbb{P}^{20}$, 
and $\check{\mathbb{P}}^{6}$ is the annihilater of $\mathbb{P}^{13}$ under the dual pairing. 
Thus they are dual to each other \cite{Rodland}. 
It was showed that they are not birational equivalent, but are derived equivalent: 
\beq
D^b(Coh(X_{1}))\cong D^b(Coh(X_{2})). 
\eeq
This pair is the first example of a derived equivalence between non-birational Calabi-Yau $3$-folds \cite{BC,Ku}. 
The field theoretic derivation of this fact by the {\it glop} (Grassmannian flop) transition was discussed in \cite{HT}. 
The pfaffians treated in this paper may also have such duality between A-model side geometries of the large and small moduli limits.

\section{Direct integration via analytic continuation}
\label{formal}
In this section, we discuss the new method for computing the B-type D-brane superpotential, proposed in \cite{FNSS}. 
We call it the direct integration via the analytic continuation. 
This method is very useful to compute the D-brane superpotential efficiently, 
so is also effective in the context of open mirror symmetry. 
By this method we can reproduce the well-known results of the open mirror symmetry for hypersurfaces and complete intersections \cite{FNSS}, 
thus refer to that paper for applications of this method to such rather simple models. 
Hereinafter we will show that this method also works very well for the pfaffian Calabi-Yau manifolds. 

\subsection{Application to the pfaffian case}
Let  us start to recall the basic procedure of this method. 
The idea is very simple: 
by introducing the new suitable coordinates and using analytic continuation, we perform the period integrals \eqref{period} directly. 
We explain this method in detail with a view to applying to the pfaffian cases. 
In the following, 
the several periods are denoted by $\Pi$ (concretely, the fundamental period, the logarithmic period, and the domainwall tension are denoted by $\Pi_0$, $\Pi_1$, and $\mathcal{T}$, respectively). 
\\
\\
$\square$Step $1$: 

First we introduce the new local coordinates of Calabi-Yau\footnote{
In \cite{FNSS}, when applying this method to the well-known models, 
we first transform the ordinary coordinates to the resolved ones, 
then we introduce the new suitable coordinates for computation of our method. 
This procedure is clear and convenient for computations of models analyzed in that paper, but in fact slightly redundant. 
As a matter of fact, we do not need to take the above two procedures, 
it is possible to take new coordinates directly, 
especially for the pfaffian Calabi-Yau manifolds we treat here. 
}. 
The pfaffian Calabi-Yau manifolds we treat in this paper have a favorable structure, 
in the sense that the hyperplanes, cutting by which we define the curve, 
are appearing explicitly in the defining equations. 
See for example \eqref{13def.eq.}. 
We introduce the new local coordinates by the following simple rule: 
\begin{itemize}
\item Take a local patch where one of hyperplanes is the form $1+x_i=0$. 
\item Introduce the new coordinates, $T$, $Y$ and $\zeta$, on two hyperplanes as follows: 
\begin{align}
&1+x_i=0 \longrightarrow 1+T=0, \\ 
&x_j+x_k=0 \longrightarrow Y^a(\zeta+\zeta^{-1})=0. 
\end{align}
\end{itemize}
We call $\zeta$ as the {\it polar} coordinate for a geometric reason, as explained in \cite{FNSS}\footnote{
Here we use a different convention from \cite{FNSS} 
and the $\zeta$-integral is performed by $\zeta=e^{i\theta}$, $\theta:0\sim 2\pi$. 
}. 
These new coordinates are useful because the $T$- and $\zeta$-part in the period integrals have the same structures for all models we treat in this paper.
The value of $a$ will be fixed later by the condition that the degrees of $Y$ in the defining equations must become $1$ or $-1$. 
See the concrete computations in section \ref{main}. 
\begin{itemize}
\item Introduce the remaining part of the new coordinates, $S$, $U$ and $W$, 
with respect to each invariant monomial under the orbifold group action, 
which can be naturally seen from the defining equations. 
\end{itemize}
For example, if the defining equations take the form 
\beq
P_a=f_a(x_i)+g_a(x_j)(x_l+x_m)+h_a(x_k), \ \ \ a=1,2,3, \label{pfa_def_eq}
\eeq
then we introduce a new coordinate, say, $S$, as follows: 
\beq
S:=\frac{f_1(x_i)}{h_1(x_k)}. \label{assign}
\eeq
Similarly, $U$ and $W$ are introduced for the other defining equations with $a=2$ and $3$. 

In fact, other assignments of new coordinates instead of $S$ are possible, for example, we can assign $1/S$, $1-S$, $S^\alpha$, $S/W$, $(1-S)/S$, and so on. 
The different choices may correspond to different expressions associated with several hypergeometric identities. 
Indeed, in the concrete computations in section \ref{main}, we perform some coordinate transformations in order to obtain the appropriate expressions of periods. 
This procedure is crucial for obtaining the correct domainwall tension as mentioned in the previous section. 

We note that, for the models except $X_{13}$, one of the defining equations is different from \eqref{pfa_def_eq}, as can be seen e.g. in \eqref{singular}. 
The brane locus is quite singular and the rule for 
introducing the local coordinates are somewhat different, as shown in \eqref{assignmentbrane}. 
\\
\\
$\square$Step $2$: 

Then we compute the period integral in the new local coordinates directly. 
Now we have introduced six new coordinates, $T$, $\zeta$, $Y$, $S$, $U$ and $W$. 
First we compute integrals of two of them by the ordinary residue theorem, picking up suitable poles. 
As a result, we obtain the following integral formula (for the $\mathbb{Z}_2$ vacua case): 
\beq
\Pi=\int \frac{C}{A-z^{1/2} B}, \label{hypersurface_like}
\eeq
where $A$, $B$ and $C$ are certain polynomials of the remaining variables. 
It is worth noting that the $\mathbb{Z}_2$ vacua structure can be explicitly seen in this formula as a power of $z$. 
($z$ is a genuine parameter of moduli, $z=\psi^{-\alpha}$, and $\alpha=7$, $10$, $9$ and $8$ for $(\check{X}_{13})_{\psi}$, $(\check{X}_5)_{\psi}$, $(\check{X}_7)_{\psi}$ and $(\check{X}_{10})_{\psi}$ respectively.) 
\\
\\
$\square$Step $3$: 

Then we obtain the factorized integral formula in the following way. 
First we expand the integral formula \eqref{hypersurface_like} with respect to the genuine moduli. 
In the pfaffian case, some of the models have the {\it large moduli}/{\it small moduli} limit points. 
Our formalism is useful for both analyses. 
For the analysis at large moduli limit, we expand \eqref{hypersurface_like} with respect to $z^{1/2}\ll 1$, 
\beq
\Pi_{\text{large}}=\int \frac{C}{A-z^{1/2} B}= \sum_{n=0}^{\infty}z^{n/2}\int \frac{B^nC}{A^{n+1}}, 
\eeq
on the contrary, for the analysis at small moduli limit, expand with respect to $z^{-1/2}\ll 1$, 
\beq
\Pi_{\text{small}}=\int \frac{C}{A-z^{1/2} B}=-z^{-1/2}\int \frac{C}{B-z^{-1/2}A}= -z^{-1/2}\sum_{n=0}^{\infty}z^{-n/2}\int\frac{A^nC}{B^{n+1}}. 
\eeq
Thus we can regard the shift in $\Theta\longrightarrow \Theta+1/2$ in \cite{Kanazawa} as the effect of the operation $\Pi_{\text{small}}\longrightarrow z^{1/2}\Pi_{\text{small}}$. 
In section \ref{main}, we shall frequently use $\Pi'$ for denoting the period integral in the small moduli region. 

It is important that for the pfaffian cases, 
one more procedure is needed to obtain the factorized form of the period inregral. 
The reason for this comes from the fact that the integrand have a factor $(a+b)^i$, where $a$ and $b$ are some monomials of the remaining variables. 
This binomial expansion leads to the additional summation with respect to $k$, which appeared in \eqref{ksum}. 
This binomial expansion should be expressed by the Gamma function because $i$ will be analytically continued to the general complex number $s$ later (see Step $4$). 
Namely, for $|b|<|a|$ and $s\in\mathbb{C}$, 
\begin{align}
(a+b)^s&=a^s\left(1+\frac{b}{a}\right)^s
=\sum_{k}\frac{\Gamma(s+1)}{\Gamma(k+1)\Gamma(s-k+1)}\left(\frac{b}{a}\right)^ka^s\nonum\\
&=\sum_{k}\frac{\Gamma(s+1)}{\Gamma(k+1)\Gamma(s-k+1)}b^ka^{s-k}, 
\end{align}
where $\sum_{k}$ means that the summation runs over the range where the Gamma functions in the formula do not possess any singularity. 
This range is determined after performing the $s$-integration (picking up poles at certain integral points). 
After this procedure, we obtain a desirable factorized integral formula. 
\\
\\
$\square$Step $4$: 

Then we perform an analytic continuation of the summation with respect to $n$ 
to the residue integral of Barnes type: 
\beq
\sum_n f_n\longrightarrow \oint\frac{ds}{2\pi i}\frac{\pi\cos(\pi s)}{\sin(\pi s)}f(s), 
\eeq
where the contour of the $s$-integration encircles all non-negative integers. 

In general, resolutions of singularities arising from the orbifolding construction is needed to analyze Calabi-Yaus and 
typically the toric method is used for obtaining new resolved coordinates 
(see e.g., \cite{MW,KW1,KS1}). 
Our coordinates introduced in Step $1$ seem not to be the resolved coordinates. 
Nevertheless, by the above analytic continuation, several integrals can be treated without suffering from the problem of singularities. 
We expect that this procedure can be regarded as the alternative procedure for resolution of singularities. 
\\
\\
$\square$Step $5$: 

The remaining problem is how to choose the contours of each integral and compute each integral via the incomplete Gamma function formula: 
\beq
\oint \frac{dv}{2\pi i}v^{\alpha-1}(1-v)^{-s-1}=
e^{-\pi i s}\frac{\sin \pi s}{\pi}\int_{0}^{1} dvv^{\alpha-1}(1-v)^{-s-1}=
-e^{-\pi i s}\frac{\Gamma(\alpha)}{\Gamma(\alpha-s)\Gamma(s+1)}, \label{gamma}
\eeq
where the contour encircles $0$ and $1$ in a counterclockwise fashion. 
From the factorized integral formula, we can obtain both the fundamental period and the domainwall tension. 
The difference between them is simply the difference between a contour and a line integral, with respect to one of local coordinates. 
It is worth noting that the structure of the $\zeta$- and $T$-integral are the same for all of the models we treat in this paper. 
After evaluating all integrals over local coordinates and performing the residue integral over $s$, 
we can obtain desirable analytic formulas of the fundamental/logarithmic periods and the domainwall tension.

For {\it the fundamental period}, 
we choose contours (closed paths) for all of variables. 
This procedure is due to the fact that the fundamental period is defined as the integral of the holomorphic $3$-form over a certain {\it $3$-cycle}, i.e., a boundary-less object. 
Since there are some possibilities of contours encircling several poles, 
we must find the suitable choice of contours which give the correct formula of the fundamental period. 
The residue integral with respect to $s$ have single poles at $s=2n$, 
and the result is the form of summation with respect to $z^{n}$. 

For {\it the domainwall tension}, 
we replace one of contour integrals with a line integral: 
\beq
\oint\longrightarrow \int_{-1}^{1}. \label{line}
\eeq
We regard the resulting integral as a $3$-chain integral, 
and intuitively the two end points of this line integral can be identified as the positions of two branes \footnote{
For all of the models except $\check{X}_{13}$, the brane is singular and its position is somewhat unclear. 
Nevertheless we can also apply this operation to those models. 
}. 
By this operation, the number of sine factors in the integrand changes and 
the pole structure of the residue integral with respect to $s$ is affected. 
It is possible to divide the domainwall tension into two parts 
\beq
\mathcal{T}=\int_{-1}^{1}(...)=\int_{0}^{1}(...)+\int_{-1}^{0}(...)=\mathcal{W}_{+}-\mathcal{W}_{-}, \label{tau}
\eeq
and we can regard two integrals as each superpotential of two branes, $\mathcal{W}_\pm$ \footnote{
Note that we use different conventions from \cite{FNSS}, for \eqref{line} and \eqref{tau}. 
Furthermore, for $\check{X}_5$, the situation is different as explained in more depth in section \ref{5}. 
}. 
After the operation \eqref{line}, the $s$-integration has double poles at $s=2n$ and single poles at $s=2n+1$, and the result is the sum of the fundamental period $\Pi_0$ and the logarithmic periods $\Pi_1$ (the summation with respect to $z^n$) and the domainwall tension $\tau$ (the summation with respect to $z^{n+1/2}$): 
\beq
\mathcal{W}_{\pm}(z)=\frac{1}{2\pi i}\frac{\Pi_1(z)}{2}\pm\frac{\Pi_0(z)}{4}\pm\frac{1}{2}\tau (z). \label{ta}
\eeq
$\mathcal{W}_+$ and $\mathcal{W}_-$ turn out to be related to each other by the following simple relation: 
\beq
\mathcal{W}_-(z^{1/2})=-\mathcal{W}_+(-z^{1/2}). \label{w-}
\eeq
Thus, we may concentrate on the evaluation of $\mathcal{W}_+$. 
$\tau$ in \eqref{ta} is nothing but the generating function of disk invariants under the A-model interpretation \eqref{A-T}. 
We use the normalization 
\beq
\sum_{d, k:\text{ odd}}\frac{2n_d^{(0,\text{real})}}{k^2}q^{kd/2}=\frac{\pi^2}{2}\frac{\tau(z)}{\Pi_0(z)}, \label{normalization}
\eeq
according to the well-known cases of hypersurfaces/complete intersections. 
In section \ref{one-loop}, we will discuss possible modifications of this overall normalization by one-loop considerations.

We note that the method via the inhomogeneous Picard-Fuchs equation \cite{MW} determines the domainwall tension up to solutions of the homogeneous equation, and such ambiguities can be determined via the monodromy arguments 
(see e.g. \cite{Wa1,KW1,KS1}). 
Our method may enable us to obtain superpotentials of both branes rather economically, although we have a subtle problem with a normalization as mentioned below.

\subsection{Normalization}
Before closing this section, we comment on certain subtleties of the normalization problem. 
We should be careful with the normalization for both the domainwall itself and the mirror relation. 
In \cite{Walcher}, it was discussed that the correct normalization of the domainwall tension is related to the order of a certain orbit of the orbifold group. 
The normalization of the mirror relation can be determined exactly
by comparison with the A-model calculation. 

In our formalism the normalization problem is very delicate since the procedure of analytic continuation may result in unclear overall factors. 
For example, we have ambiguities of the form $e^{2\pi is}$ resulting from the choice of direction of some contours. 
These ambiguities do not affect the final result since in the end the values of $s$ must be integer. 
However it is possible that worse ambiguities appear.

Furthermore, for the domainwall tension, we have the additional ambiguity of an overall normalization due to contributions from some local patches. 
The problem is that the holomorphic period \eqref{hol3form} is defined on the affine patch $\{P_{\nu_0\nu_1\nu_2}\neq 0\}$. 
Furthermore, to cover the entire brane, we need some local patches. 
The correct normalization of the B-brane superpotential can be obtained by taking into account the contributions from all patches \cite{FNSS}, and 
in pfaffian cases, it is not clear how to consider several contributions properly.

Since we do not have the A-model pictures and calculations so far, instead of computing the correct normalization, 
we put the suitable normalization factor by hand 
in this paper. 
We try to fix this ambiguity by imposing the integrality of real BPS numbers of both tree and one-loop level in section \ref{one-loop}.

\section{Disc instantons of Pfaffian Calabi-Yau $3$-folds}
\label{main}
In this section, we carry out the concrete computations of the direct integration via the analytic continuation discussed in the previous section. 

\subsection{Mirror of degree $13$}
The large moduli limit of $\check{X}_{13}$ is the model No.99 and the small moduli limit of this is the model No.225 in \cite{data}. 

By choosing a suitable permutation $\nu$ in \eqref{period}, 
the holomorphic $3$-form can be expressed by the following equations: 
\beq
\Pi=\frac{13}{(2\pi i)^6}\int\frac{P_{134}}{P_{1}P_{3}P_{4}}\omega_0, 
\eeq
where 
\begin{align}
&P_{1}=x_0x_3x_4-\frac{1}{\psi}x_5x_6(x_5+x_6)-\frac{1}{\psi^2}x_1x_2^2, \label{13def.eq.}\\
&P_{3}=x_1x_5x_6-\frac{1}{\psi}x_0^3-\frac{1}{\psi^2}x_2^2(x_3+x_4), \\
&P_{4}=x_2x_5x_6-\frac{1}{\psi}x_3x_4(x_3+x_4)-\frac{1}{\psi^2}x_0^2x_1, \\
&P_{134}=x_5x_6. 
\end{align}
As noted previously, we choose the B-brane (holomorphic curve) as follows: 
\begin{align}
&x_3+x_4=x_5+x_6=0, \\
&x_0x_3^2+\frac{1}{\psi^2}x_1x_2^2=0, \ \ 
x_1x_5^2+\frac{1}{\psi}x_0^3=0, \ \ 
x_2x_5^2+\frac{1}{\psi^2}x_0^2x_1=0. 
\end{align}
In the following computation, we take a local patch $x_3=1$. 
The computations in other patches are exactly the same and their contributions merely change the numerical overall factor of the final result. 
The overall factor of the domainwall tension will be discussed later by considering the one-loop amplitude. 
The other models are treated in a similar way. 

Following the general theory, we introduce new coordinates. 
$T$, $Y$ and $\zeta$ are introduced for hyperplanes as explained in section \ref{formal}, 
and $W$, $S$ and $U$ are introduced as below: 
\begin{align}
\frac{x_0x_3x_4}{x_1x_2^2}=\frac{1}{\psi^2}\frac{1}{W^{2}}, \ \
\frac{x_1x_5x_6}{x_0^3}=\frac{1}{\psi}S, \ \
\frac{x_2x_5x_6}{x_0^2x_1}=\frac{1}{\psi^2}U. \label{13law}
\end{align}
The power of $W$ is chosen for later convenience. 
Then we obtain the following transformation law: 
\begin{align}
&x_0=\psi^{3/4}Y^{5a/6}T^{1/12}S^{-1/4}U^{-1/6}W^{1/6},\nonum\\
&x_1=\psi^{5/4}Y^{3a/6}T^{1/4}S^{1/4}U^{-1/2}W^{1/2},\nonum\\
&x_2=\psi^{3/4}Y^{a/6}T^{5/12}S^{-1/4}U^{1/6}W^{5/6},\nonum\\ 
&x_3=1, \ \ x_4=T, \ \ x_5=Y^{a}\zeta, \ \ x_6=Y^{a}\zeta^{-1}. 
\end{align}
The defining equations in the new coordinates becomes ($a$ is fixed to $6/13$) 
\begin{align}
P_1&=\psi^{3/4}Y^{5/13}T^{13/12}S^{-1/4}U^{-1/6}W^{1/6}[1-W^{2}-\psi^{-7/4}(\zeta+\zeta^{-1})T^{-13/12}YS^{1/4}U^{1/6}W^{-1/6}], \\
P_3&=\psi^{5/4}Y^{15/13}T^{1/4}S^{-3/4}U^{-1/2}W^{1/2}[S-1-\psi^{-7/4}(1+T)T^{7/12}Y^{-1}S^{1/4}U^{5/6}W^{7/6}], \\
P_4&=\psi^{3/4}YT^{5/12}S^{-1/4}U^{-5/6}W^{5/6}[U-1-\psi^{-7/4}(1+T)T^{7/12}Y^{-1}S^{1/4}U^{5/6}W^{-5/6}]. 
\end{align}
The defining equations of the brane in the new coordinates are 
\bea
1+T=0, \ \ Y^{a}(\zeta+\zeta^{-1})=0, \ \ S=1, \ \ U=1, \ \ W=\pm 1. 
\ena
Thus the curve can be regarded as two families with respect to the discrete open moduli $\pm$. 

Now, we introduce a genuine moduli $z=\psi^{-7}$. 
Then the period integral is 
\begin{align}
\Pi=\frac{2}{(2\pi i)^6}\int \frac{dYdTd\zeta dWdUdS}{YT\zeta W}
&\frac{1}{[1-W^{2}-z^{1/4}(\zeta+\zeta^{-1})T^{-13/12}YS^{1/4}U^{1/6}W^{-1/6}]}\nonum\\
\times &\frac{1}{[S-1-z^{1/4}(1+T)T^{7/12}Y^{-1}S^{1/4}U^{5/6}W^{7/6}]}\nonum\\
\times &\frac{1}{[U-1-z^{1/4}(1+T)T^{7/12}Y^{-1}S^{1/4}U^{5/6}W^{-5/6}]}. 
\end{align}
First we perform the $Y$-integration by picking up the pole at 
$Y^{-1}=\frac{U-1}{z^{1/4}T^{7/12}(1+T)S^{1/4}U^{5/6}W^{-5/6}}$. 
\begin{align}
\Pi=\frac{2}{(2\pi i)^5}\int\frac{dTd\zeta dWdUdS}{T\zeta W}&\frac{1}{[(1-W^{2})(U-1)-z^{1/4}(T^{-1/2}+T^{1/2})(\zeta+\zeta^{-1})S^{1/2}UW^{-1}]}\nonum\\
\times &\frac{1}{[S-1-W^{2}(U-1)]}. 
\end{align}
The $U$-integration is performed by picking up the pole at $U=1+W^{-2}(S-1)$, 
\begin{align}
\Pi=\frac{2}{(2\pi i)^4}\int\frac{dTd\zeta dWdS}{T\zeta W}\frac{1}{[(S-1)(1-W^{2})-z^{1/2}(T^{1/2}+T^{-1/2})(\zeta+\zeta^{-1})S^{1/2}W^{-1}(W^{2}-1-S)]}. 
\end{align}
In the following, we concentrate on the evaluation of $\mathcal{W}_+$. 
$\mathcal{W}_-$ can be obtained easily using the relation \eqref{w-}. 
We introduce a new coordinate $X=W^{2}$ and the following notation: 
\beq
A(T,\zeta):=(T^{1/2}+T^{-1/2})(\zeta+\zeta^{-1})\label{A}. 
\eeq
Then, the period integral becomes 
\begin{align}
\Pi=\frac{1}{(2\pi i)^4}\int\frac{dTd\zeta dXdS}{T\zeta X}
\frac{1}{[(S-1)(1-X)-z^{1/2}A(T,\zeta)S^{1/2}X^{-1/2}(X+(S-1))]}. \label{13period}
\end{align}
This is the basic formula for both large and small moduli analyses.

\subsubsection{Large moduli limit}
First we consider the large moduli limit, $z^{1/2}\ll 1$. 
As noted previously this is the model No.99 in \cite{data}. 
By expanding around the large moduli limit point, 
the period integral becomes 
\begin{align}
\Pi=\sum_{n=0}^\infty z^{n/2}\int \frac{dTd\zeta}{(2\pi i)^2T\zeta}A(T,\zeta)^{n}
\int \frac{dXdS}{(2\pi i)^2}\frac{S^{n/2}X^{-n/2-1}(X+(S-1))^{n}}{(S-1)^{n+1}(1-X)^{n+1}}. \label{13binomial2}
\end{align}
Using the following binomial expansion: 
\beq
(X+(S-1))^n=\sum_{k}\frac{\Gamma(n+1)}{\Gamma(k+1)\Gamma(n-k+1)}(S-1)^kX^{n-k}, \label{13binomial1}
\eeq
we obtain the following factorized integral: 
\begin{align}
\Pi= &\sum_{n=0}^\infty z^{n/2}\oint \frac{d\zeta}{2\pi i}\frac{(\zeta^2+1)^n}{\zeta^{n+1}}\oint \frac{dT}{2\pi i}\frac{(T+1)^n}{T^{n/2+1}}
\sum_{k=0}^n\frac{\Gamma(n+1)}{\Gamma(k+1)\Gamma(n-k+1)}\nonum\\
&\times\oint \frac{dX}{2\pi i}X^{n/2-k-1}(1-X)^{-n-1}
\oint \frac{dS}{2\pi i}S^{n/2}(1-S)^{-n+k-1}(-1)^{-n+k-1}. 
\end{align}
Then we perform the analytic continuation of the summation to the Barnes-type integral formula: 
\begin{align}
\Pi= &\oint \frac{ds}{2\pi i}\frac{\pi\cos(\pi s)}{\sin(\pi s)}z^{s/2}
\oint \frac{dT}{2\pi i}\frac{(T+1)^s}{T^{s/2+1}}
\oint \frac{d\zeta}{2\pi i}\frac{(\zeta^2+1)^s}{\zeta^{s+1}}
\sum_{k}\frac{\Gamma(s+1)(-1)^{-s+k+1}}{\Gamma(s-k+1)\Gamma(k+1)}\nonum\\
&\times\oint \frac{dS}{2\pi i}S^{s/2}(1-S)^{-s+k-1}
\oint \frac{dX}{2\pi i}X^{s/2-k-1}(1-X)^{-s-1}. 
\end{align}
Each integral in the above formula can be evaluated as follows: 
The $\zeta$- and $T$-integral can be evaluated as 
\begin{align}
&\oint \frac{d\zeta}{2\pi i}\frac{(\zeta^2+1)^s}{\zeta^{s+1}}=
\oint \frac{d\zeta}{2\pi i}\zeta^{-s-1}(1+\zeta^2)^s=e^{-\pi i s/2}\cos\left(\frac{\pi s}{2}\right)\frac{\Gamma(s+1)}{\Gamma(\frac{s}{2}+1)^2}, \label{zeta}\\
&\oint \frac{dT}{2\pi i}\frac{(T+1)^s}{T^{s/2+1}}
=\oint \frac{dT}{2\pi i}T^{-s/2-1}(1+T)^s
=\frac{\Gamma(s+1)}{\Gamma(\frac{s}{2}+1)^2}. \label{T}
\end{align}
These integrals are common for all the models in this paper. 
The $S$-integral gives 
\begin{align}
\oint \frac{dS}{2\pi i}S^{s/2}(1-S)^{-s+k-1}=-e^{-\pi i(s-k)}\frac{\Gamma(s/2+1)}{\Gamma(k-s/2+1)\Gamma(s-k+1)}. 
\end{align}
For the $X$-integral, we choose a contour integral for the fundamental period and a line integral for the domainwall tension. 

First we consider the fundamental period. 
In this case we have 
\begin{align}
\oint \frac{dX}{2\pi i}X^{s/2-k-1}(1-X)^{-s-1}=-e^{-\pi i s}\frac{\Gamma(s/2-k)}{\Gamma(-s/2-k)\Gamma(s+1)}. 
\end{align}

By collecting all results and performing the $s$-integration, 
we have simple poles at $s=2n$ and the period integral becomes 
\begin{align}
\Pi_0=&
\oint \frac{ds}{2\pi i}\frac{\pi\cos(\pi s)\cos(\pi s/2)}{\sin(\pi s)}e^{-\pi is/2}(-1)^sz^{s/2}
\frac{\Gamma(s+1)^2}{\Gamma(\frac{s}{2}+1)^4}
\sum_{k} \frac{\Gamma(s+1)}{\Gamma(k+1)\Gamma(s-k+1)}\nonum\\
&\times \frac{\Gamma(s/2-k)}{\Gamma(s+1)\Gamma(-s/2-k)}
\frac{\Gamma(s/2+1)}{\Gamma(s-k+1)\Gamma(-s/2+k+1)}\nonumber\\
=&
\sum_{n=0}^\infty z^{n}
{\begin{pmatrix} 2n \\ n \end{pmatrix}}^2
\sum_{k=n}^{2n}
\frac{\Gamma(n+k+1)\Gamma(n+1)}{\Gamma(k+1)\Gamma(2n-k+1)^2\Gamma(k-n+1)^2}. 
\end{align}
In the above formula, the summation with respect to $k$ runs from $n$ to $2n$. 
After changing the summation variable to $l=k-n$, the fundamental period becomes 
\begin{align}
\Pi_0= &\sum_{n=0}^\infty z^{n}
{\begin{pmatrix} 2n \\ n \end{pmatrix}}^2
\sum_{l=0}^{n} \frac{\Gamma(2n+l+1)\Gamma(n+1)}{\Gamma(n+l+1)\Gamma(n-l+1)^2\Gamma(l+1)^2}\nonum\\
= &\sum_{n=0}^\infty z^{n}
{\begin{pmatrix} 2n \\ n \end{pmatrix}}^2
\sum_{l=0}^{n}
{\begin{pmatrix} 2n+l \\ n \end{pmatrix}}
{\begin{pmatrix} n \\ l \end{pmatrix}}^2. \label{13funda}
\end{align}
This is exactly the same as the formula in \cite{data} and \cite{Kanazawa}.

Then we consider the domainwall tension (precisely, one of superpotentials, $\mathcal{W}_+$). 
We replace one of contour integrals (with respect to $X$) with a line integral. 
As noted previously, $\mathcal{W}_-$ can be obtained easily by transforming $z^{1/2}\longrightarrow -z^{1/2}$. 
This is also true for the other models except $\check{X}_5$. 

The line integral of $X$ is evaluated as 
\bea
\int_{0}^{1}dX X^{s/2-k-1}(1-X)^{-s-1}
=\frac{\Gamma(s/2-k)\Gamma(-s)}{\Gamma(-s/2-k)}. 
\ena

Thus we obtain 
\begin{align}
\mathcal{W}_+
=&\frac{1}{2\pi i}\int\frac{ds}{2\pi i}\frac{\pi \cos(\pi s)\cos (\pi s/2)}{\sin^2(\pi s)}z^{s/2}\frac{\Gamma(s+1)^2}{\Gamma(s/2+1)^4}
\sum_{k} \frac{\Gamma(s+1)}{\Gamma(k+1)\Gamma(s-k+1)}\nonum\\
&\times \frac{\Gamma(-s)\Gamma(s/2-k)}{\Gamma(-s/2-k)}
\frac{\Gamma(s/2+1)}{\Gamma(s-k+1)\Gamma(-s/2+k+1)}. 
\end{align}
This has single poles at $s=2n+1$ and double poles at $s=2n$. 
Thus, the result is $\mathcal{W}_+=\frac{1}{4\pi i}\Pi_1+\frac{1}{4}\Pi_0+\frac{1}{2}\tau$, where 
$\Pi_0$ is the fundamental period given in \eqref{13funda}, 
$\Pi_1$ is the logarithmic period given by 
\begin{align}
\Pi_1= &\Pi_0\log z+\sum_{n=0}^{\infty}z^n
{\begin{pmatrix} 2n \\ n \end{pmatrix}}^2
\sum_{k=0}^n\begin{pmatrix} 2n+k \\ n \end{pmatrix}{\begin{pmatrix} n \\ k \end{pmatrix}}^2\nonum\\
&\times[4 \Psi(2n+1)-3\Psi(n+1)+2\Psi(2n+k+1)-\Psi(n+k+1)-2\Psi(n-k+1)], 
\end{align}
and $\tau$ is given by 
\begin{align}
\tau=\sum_{n=0}^{\infty}z^{n+1/2}\frac{\Gamma(2n+2)^2}{\Gamma(n+\frac{3}{2})^4}\sum_{k=0}^{2n+1}
\frac{\Gamma(n+k+\frac{3}{2})\Gamma(n+\frac{3}{2})}{\Gamma(k-n+\frac{1}{2})^2\Gamma(2n+2-k)^2\Gamma(k+1)}. \label{13domain}
\end{align}

By the use of the mirror map, 
the formula \eqref{13domain} with a normalization \eqref{normalization} 
gives the following integral real BPS invariants (up to overall normalization) of low degrees: 
\beq
n_{1}^{(0,\text{real})}=7, \ 
n_{3}^{(0,\text{real})}=35, \ 
n_{5}^{(0,\text{real})}=2564, \ 
n_{7}^{(0,\text{real})}=270402, \ 
n_{9}^{(0,\text{real})}=32866812, \ .... 
\eeq
The real BPS invariants of higher degrees are listed in appendix \ref{invariants}.

\subsubsection{Small moduli limit}
This is the model No.225 in \cite{data}. 
Let us restart with the formula \eqref{13period}: 
\begin{align}
\Pi=\frac{1}{(2\pi i)^4}\int\frac{dTd\zeta dXdS}{T\zeta X}
\frac{1}{[(S-1)(1-X)-z^{1/2}A(T,\zeta)S^{1/2}X^{-1/2}(X+S-1)]}. \label{a}
\end{align}

Now we consider small moduli limit, $z^{-1/2}\ll 1$, and 
rewrite \eqref{a} as 
\begin{align}
\Pi= &-z^{-1/2}\int\frac{dT\zeta dXdS}{(2\pi i)^4T\zeta X}\frac{1}{[A(T,\zeta)S^{1/2}X^{-1/2}(X+S-1)+z^{-1/2}(1-X)(1-S)]}. 
\end{align}

The shift $\Theta\longrightarrow \Theta+1/2$ noted in \cite{Kanazawa} can be naturally explained as the following gauge transformation of the period, 
\beq
\Pi\longrightarrow \Pi'=z^{1/2}\Pi. \label{shift1}
\eeq

Now consider the following transformations $(T,\zeta)\longrightarrow(T',\zeta')$: 
\beq
\frac{1}{2}(T^{1/2}+T^{-1/2})=\frac{1}{\frac{1}{2}({T'}^{1/2}+{T'}^{-1/2})}, \ \ 
\frac{1}{2}(\zeta+\zeta^{-1})=\frac{1}{\frac{1}{2}(\zeta'+{\zeta'}^{-1})}. \label{Tzeta}
\eeq
Then, 
\beq
\frac{dTd\zeta}{T\zeta}=-\frac{1}{\frac{1}{2^4}({T'}^{1/2}+{T'}^{-1/2})(\zeta'+{\zeta'}^{-1})}\frac{dT'd\zeta'}{T'\zeta'}
=-2^{4}{A'(T',\zeta')}^{-1}\frac{dT'd\zeta'}{T'\zeta'}, 
\eeq
where $A'(T',\zeta')=({T'}^{1/2}+{T'}^{-1/2})(\zeta'+{\zeta'}^{-1})=2^4A(T,\zeta)^{-1}$. 
In the following, we drop the prime symbol $'$ for simplicity. 

As a result, the period integral becomes 
\begin{align}
\Pi'=z^{1/2}\Pi=\int\frac{dTd\zeta dXdS}{(2\pi i)^4T\zeta X}\frac{1}{[S^{1/2}X^{-1/2}(X+S-1)+2^{-4}z^{-1/2}A(T,\zeta)(1-X)(1-S)]}. 
\end{align}
To obtain the appropriate form of the periods, 
we perform the coordinate transformations, 
$S\longrightarrow 1/S$, 
$X\longrightarrow \frac{S-1}{S}X$ and 
$S\longrightarrow 1-S$, in turn. 
Then we obtain 
\begin{align}
\Pi'=\int\frac{dTd\zeta dXdS}{(2\pi i)^4T\zeta X^{1/2}S^{1/2}}
\frac{1}{[(1-S)(1-X)+2^{-4}z^{-1/2}A(T,\zeta)S^{1/2}X^{1/2}((1-S)(1-X)+X)]}. 
\end{align}
Expanding this around the small moduli limit point, we find 
\begin{align}
\Pi'=&\sum_{n=0}^{\infty}(-2^{-4}z^{-1/2})^n\int\frac{dTd\zeta}{(2\pi i)^2T\zeta}A(T,\zeta)^{n}\int\frac{dXdS}{(2\pi i)^2}
\frac{S^{n/2-1/2}X^{n/2-1/2}((1-S)(1-X)+X)^n}{(1-S)^{n+1}(1-X)^{n+1}}. 
\end{align}
Under the following binomial expansion: 
\begin{align}
((1-S)(1-X)+X)^{n}=
\sum_{k}\frac{\Gamma(n+1)}{\Gamma(n-k+1)\Gamma(k+1)}X^{n-k}(1-S)^k(1-X)^k, 
\end{align}
and performing the analytic continuation, we obtain the factorized integral 
\begin{align}
\Pi'=&\int\frac{ds}{2\pi i}\frac{\pi\cos(\pi s)}{\sin(\pi s)}e^{-\pi is/2}(-1)^s(-2^{-4}z^{-1/2})^s
\int\frac{dTd\zeta}{(2\pi i)^2T\zeta}A(T,\zeta)^s
\sum_{k}\frac{\Gamma(s+1)}{\Gamma(s-k+1)\Gamma(k+1)}\nonum\\
&\times
\int \frac{dX}{2\pi i}X^{3s/2-k-1/2}(1-X)^{-s+k-1}
\int \frac{dS}{2\pi i}S^{s/2-1/2}(1-S)^{-s+k-1}. 
\end{align}

First, let us consider the fundamental period, $\Pi_0'$. 
The $T$- and $\zeta$-integral are the same as \eqref{T} and \eqref{zeta}, respectively. 
The other parts can be also evaluated easily by using \eqref{gamma}. 
Then we have simple poles at $s=2n$. 
By performing the residue integral with respect to $s$, 
and by introducing the genuine moduli $x=-(2^{16}z)^{-1}$, 
the fundamental period in the small moduli region becomes 
\bea
\Pi'_0=\sum_{n=0}^{\infty}x^n2^{8n}
\begin{pmatrix} 2n \\ n \end{pmatrix}^2
\sum_{k=0}^{2n}\frac{\Gamma(2n+1)\Gamma(3n-k+1/2)}{\Gamma(k+1)\Gamma(2n-k+1)^3\Gamma(-n+k+1/2)}. \label{13smallfunda}
\ena

At first sight, the above formula seems strange since the arguments of the Gamma functions are half integers although this is the fundamental period. 
Moreover, we have extra $2^{8n}$ factor which looks like redundant. 
However, it is possible to bring this extra $2^{8n}$ factor into the Gamma functions and solve these problems simultaneously. 
In \cite{Kanazawa}, 
it was observed that $x=-(2^{16}z)^{-1}$ is the genuine moduli parameter, 
although the meaning of the factor $2^{16}$ is unclear. 
It is interesting that by our method, we can naturally obtain the good moduli parameter $x$ as above. 
The meaning of the shift $\Theta\longrightarrow \Theta+1/2$ noted in \cite{Kanazawa} can be also understood naturally in \eqref{shift1}. 
These arguments are also valid for the small moduli analyses of the other models.

Now, we consider the domainwall tension (precisely $\mathcal{W}_+'$). 
We replace the contour integrals with respect to $X$ with the line integral. 
As a result, the period integral have simple poles at $s=2n+1$ and double poles at $s=2n$. 
Thus we obtain 
$\mathcal{W}_{+}'=\frac{1}{4\pi i}\Pi_{1}'+\frac{1}{4}\Pi_{0}'+\frac{1}{2}\tau'$, 
where $\Pi_0'$ is the fundamental period given in \eqref{13smallfunda}, 
$\Pi_1'$ is the logarithmic period given by 
\begin{align}
\Pi_1'=&\Pi_0'\log z+\sum_{n=0}^{\infty}x^n2^{8n}
\begin{pmatrix} 2n \\ n \end{pmatrix}^2
\sum_{k=0}^{2n}\frac{\Gamma(2n+1)\Gamma(3n-k+1/2)}{\Gamma(-n+k+1/2)\Gamma(2n-k+1)^3\Gamma(k+1)}\nonum\\
&\times[6\Psi(2n+1)-4\Psi(n+1)+3\Psi(3n-k+1/2)-6\Psi(2n-k+1)\nonum\\
&+\Psi(-n+k+1/2)+4\Psi(1)-4\Psi(1/2)], 
\end{align}
and $\tau'$ is given by 
\beq
\tau'=\sum_{n=0}^{\infty}x^{n+1/2}2^{8n+4}\frac{\Gamma(2n+2)^2}{\Gamma(n+3/2)^4}\sum_{k=n+1}^{2n+1}
\frac{\Gamma(2n+2)\Gamma(3n-k+2)}{\Gamma(-n+k)\Gamma(2n-k+2)^3\Gamma(k+1)}. 
\eeq

By this formula, we find the real BPS invariants (up to overall normalization) of low degrees: 
\beq
n_{1}^{(0,\text{real})}=64, \ 
n_{3}^{(0,\text{real})}=47680, \ 
n_{5}^{(0,\text{real})}=2553150912, \ 
n_{7}^{(0,\text{real})}=34649296391104, 
.... 
\eeq
The real BPS invariants of higher degrees are listed in appendix \ref{invariants}. 
\\

Next we consider the other models, $\check{X}_5$, $\check{X}_7$ and $\check{X}_{10}$. 
Their structure is quite similar to the one of $\check{X}_{13}$. 
However, there are some important differences. 
First, the {\it hyperplanes} of $\check{X}_5$ by which we define the brane have the degree $2$-form, $x_i^2+x_j^2=0$. 
So we need to modify the local parametrization of $T$ and $\zeta$. 
Second, for all of these three models, 
some of the local defining equations have different structures in a way that the brane equations obtained after cutting by two hyperplanes are quite singular and 
the positions of the branes are somewhat uncertain. 
Third, the most important difference, 
$\check{X}_5$ and $\check{X}_{10}$ have two special points (the large and small moduli limits) in their moduli space, just like $\check{X}_{13}$, but $\check{X}_7$ have only one special point (the large moduli limit) in its moduli space. 
Furthermore, $\check{X}_5$ is {\it self-dual}, namely, the Calabi-Yaus realized in the large/small moduli limits are in fact the same one.

\subsection{Mirror of degree $5$}\label{5}
The phaffian Calabi-Yau $\check{X}_{5}$ corresponds to the model No.302 in \cite{data} and has the so-called self-dual structure, as stressed previously. 
We can confirm this feature by explicit computations, 
at the level of the integral formula. 

We choose $P_1$, $P_2$ and $P_3$ for constructing holomorphic $3$-form: 
\begin{align}
&P_1=-x_0x_1x_6+\frac{1}{\psi}x_4^2+\frac{1}{\psi^2}(x_2^2+x_3^2)x_5, \\
&P_2=-x_0x_1x_2x_3+\frac{1}{\psi}x_5^2+\frac{1}{\psi^2}x_4x_6, \label{singular}\\
&P_3=-x_2x_3x_4+\frac{1}{\psi}x_6^2+\frac{1}{\psi^2}(x_0^2+x_1^2)x_5, \\
&P_{123}=\frac{1}{\psi}x_5. 
\end{align}
As noted previously, 
the {\it hyperplanes} are the following degree $2$ forms: 
\beq
x_0^2+x_1^2=0, \ \ x_2^2+x_3^2=0, 
\eeq
and the defining equations of the brane are 
\begin{align}
x_0x_1x_6-\frac{1}{\psi}x_4^2=0, \ \ 
x_0x_1x_2x_3-\frac{1}{\psi}x_5^2-\frac{1}{\psi^2}x_4x_6=0, \ \ 
x_2x_3x_4-\frac{1}{\psi}x_6^2=0. 
\end{align}
By introducing new coordinates as follows: 
\beq
\frac{x_0x_1x_6}{x_4^2}=\frac{1}{\psi}S, \ \ 
\frac{x_2x_3x_4}{x_6^2}=\frac{1}{\psi}X, \ \ 
\frac{x_4x_6}{x_5^2}=\psi U, \label{assignmentbrane}
\eeq
we have the following transformation law\footnote{
The powers of $T$, $\zeta$ and $Y$ for $\check{X}_{5}$ are different from those for the other models because the degree of the {\it hyperplanes} is different. 
Thus, it is natural that the $\zeta$-integration is performed by $\zeta=e^{i\theta}$, $\theta:0\sim 4\pi$, in this case. 
}: 
\begin{align}
&x_0=1, \ \ x_1=T^{1/2}, \ \ x_2=Y^{a/2}\zeta^{1/2}, \ \ x_3=Y^{a/2}\zeta^{-1/2}, \nonum\\
&x_4=\psi T^{1/3}Y^{a/3}S^{-2/3}X^{-1/3}, \nonum\\
&x_5=\psi^{1/2} T^{1/4}Y^{a/2}S^{-1/2}X^{-1/2}U^{-1/2}, \nonum\\
&x_6=\psi T^{1/6}Y^{2a/3}S^{-1/3}X^{-2/3}. 
\end{align}
Then, the defining equations in the new coordinates are ($a$ is fixed to $6/5$) 
\begin{align}
&P_1=\psi T^{2/3}Y^{4/5}S^{-4/3}X^{-2/3}
[1-S+\psi^{-5/2}(\zeta+\zeta^{-1})T^{-5/12}YS^{5/6}X^{1/6}U^{-1/2}], \nonum\\
&P_2=T^{1/2}Y^{6/5}S^{-1}X^{-1}U^{-1}[(1-SX)U+1], \nonum\\
&P_3=\psi Y^{8/5}T^{1/3}S^{-2/3}X^{-4/3}
[1-X+\psi^{-5/2}(1+T)T^{-1/12}Y^{-1}S^{1/6}X^{5/6}U^{-1/2}]. 
\end{align}
The defining equations of the brane in the new coordinates are 
\bea
1+T=0, \ \ Y^{6/5}(\zeta+\zeta^{-1})=0, \ \ S=1, \ \ (1-SX)U+1=0, \ \ X=1. 
\ena
This is interpreted as a single brane and the other brane of $\mathbb{Z}_2$ vacua will be discussed later. 

Now we introduce a genuine moduli $z=\psi^{-10}$. 
Then the period integral is 
\begin{align}
\Pi=&\frac{10}{(2\pi i)^6}
\int\frac{P_{123}}{P_1P_2P_3}\omega\nonum\\
=&\frac{1}{2}\frac{1}{(2\pi i)^6}\int\frac{dYdTd\zeta dU}{YT\zeta U}dSdX\frac{1}{[1-S+z^{1/4}(\zeta+\zeta^{-1})T^{-5/12}YS^{5/6}X^{1/6}U^{-1/2}]}\nonum\\
&\times\frac{1}{[(1-SX)U+1]}\frac{1}{[1-X+z^{1/4}(1+T)T^{-1/12}Y^{-1}S^{1/6}X^{5/6}U^{-1/2}]}. 
\end{align}
First we perform the $Y$-integration by picking up the pole at 
$Y=\frac{S-1}{z^{1/4}(\zeta+\zeta^{-1})T^{-12/5}S^{5/6}X^{1/6}U^{-1/2}}$, 
and the integral becomes 
\begin{align}
\Pi=\frac{1}{2}\frac{1}{(2\pi i)^5}\int\frac{dTd\zeta dU}{T\zeta U}dSdX\frac{1}{[(1-SX)U+1]}
\frac{1}{[(1-X)(S-1)-z^{1/2}A(T,\zeta) SXU^{-1}]}. 
\end{align}
Here we use the notation \eqref{A}. 
The $U$-integration is evaluated by picking up the pole at 
$U=\frac{1}{SX-1}$, 
and we find 
\begin{align}
\Pi=&\frac{1}{2}\frac{1}{(2\pi i)^4}\int\frac{dTd\zeta}{T\zeta}dSdX\frac{1}{[(1-X)(1-S)-z^{1/2}A(T,\zeta)SX(SX-1)]}. \label{5period}
\end{align}

\subsubsection{Large moduli limit}
First we consider the large moduli limit, $z^{1/2}\ll 1$. 
Then, expanding \eqref{5period} around the large moduli limit point, we have 
\begin{align}
\Pi=\frac{1}{2}\sum_{n=0}^{\infty}z^{n/2}\int\frac{dTd\zeta}{(2\pi i)^2T\zeta}A(T,\zeta)^{n}\int\frac{dSdX}{(2\pi i)^2}\frac{S^nX^n(SX-1)^n}{(1-X)^{n+1}(1-S)^{n+1}}. 
\end{align}
Under the following binomial expansion formula: 
\beq
(SX-1)^{n}=\sum_{k}\frac{\Gamma(n+1)}{\Gamma(n-k+1)\Gamma(k+1)}S^{n-k}X^{n-k}(-1)^{k}, 
\eeq
and performing the analytic continuation, we obtain the factorized integral 
\begin{align}
\Pi=&\frac{1}{2}\int \frac{ds}{2\pi i}\frac{\pi\cos(\pi s)}{\sin(\pi s)}z^{s/2}\int\frac{dTd\zeta}{(2\pi i)^2T\zeta}A(T,\zeta)^{s}
\sum_{k}\frac{\Gamma(s+1)(-1)^{k}}{\Gamma(s-k+1)\Gamma(k+1)}\nonum\\
&\times
\int\frac{dS}{2\pi i}S^{2s-k}(1-S)^{-s-1}
\int\frac{dX}{2\pi i}X^{2s-k}(1-X)^{-s-1}. 
\end{align}
The $T$-integral is the same as \eqref{T}. 
The $\zeta$-integral is $2\times$\eqref{zeta} (according to the fact noted in the footnote in the previous page), and the pre-factor $1/2$ in the above formula is canceled. 
The $S$- and $X$-integral can be evaluated by using \eqref{gamma}. 

Then, collecting all integrals and performing the $s$-integral, 
we have poles at $s=2n$ and the fundamental period leads to 
\begin{align}
\Pi_0
&=\sum_{n=0}^{\infty}z^n
\begin{pmatrix} 2n \\ n \end{pmatrix}^2
\sum_{k=0}^{2n}(-1)^k
\frac{\Gamma(4n-k+1)^2}{\Gamma(2n-k+1)^3\Gamma(2n+1)\Gamma(k+1)}\nonum\\
&=\sum_{n=0}^\infty z^{n}
\begin{pmatrix} 2n \\ n \end{pmatrix}^2
\sum_{k=0}^{2n}(-1)^k
\begin{pmatrix} 2n \\ k \end{pmatrix}
\begin{pmatrix} 4n-k \\ n-k \end{pmatrix}^2. \label{5funda}
\end{align}
Although the above formula seem to be different from the one obtained in \cite{Kanazawa}, 
it is possible to prove that they are equal to each other by the use of hypergeometric identities \footnote{
As a matter of fact, in \cite{binomial}, there are many lists of some different looking formulas of periods which are in fact the same. 
}. 
Such is the case with the period formulas of the other models $X_{7}$ and $X_{10}$ discussed later.

For the domainwall tension (precisely $\mathcal{W}_+$), 
by replacing the contour integral with respect to $X$ with the line integral, 
we find simple poles at $s=2n+1$ and double poles at $s=2n$. 
As a result, we obtain $\mathcal{W}_+=\frac{1}{4\pi i}\Pi_1+\frac{1}{4}\Pi_0+\frac{1}{2}\tau$, 
where $\Pi_0$ is the fundamental period given in \eqref{5funda}, 
$\Pi_1$ is the logarithmic period given by 
\begin{align}
\Pi_1=&\Pi_0\log z+\sum_{n=0}^{\infty}z^n
\begin{pmatrix} 2n \\ n \end{pmatrix}^2
\sum_{k=0}^{2n}(-1)^k\frac{\Gamma(4n-k+1)^2}{\Gamma(2n-k+1)^3\Gamma(2n+1)\Gamma(k+1)}\nonum\\
&\times [2\Psi(2n+1)-4\Psi(n+1)+8\Psi(4n-k+1)-6\Psi(2n-k+1)], 
\end{align}
and $\tau$ is given by 
\beq
\tau=\sum_{n=0}^{\infty}z^{n+1/2}\frac{\Gamma(2n+2)^2}{\Gamma(n+3/2)^4}\sum_{k=0}^{2n+1}(-1)^k
\frac{\Gamma(4n-k+3)^2}{\Gamma(2n-k+2)^3\Gamma(2n+2)\Gamma(k+1)}. \label{5T}
\eeq

It should be noted that the relation \eqref{w-} does not hold in this case, 
as can be seen from \eqref{5period}. 
We consider in the following way. 
The hyperplanes in this model are precisely expressed by $\{x_0+\alpha x_1=x_2+\beta x_3=0\}$ $(\alpha^2=\beta^2=-1)$. 
The superpotential obtained in the above calculation (it is denoted by $\mathcal{W}_+$) is that of the brane with $\alpha=+i$ and $\beta=+i$. 
Now we consider another brane with $\alpha=+i$ and $\beta=-i$. 
Then the factor $A(T,\zeta)$ in \eqref{5period} has a minus sign. 
This is interpreted as the superpotential of another brane, $\mathcal{W}_-$, which is defined by another set of the hyperplanes. 
Similar structures are observed for models treated in \cite{Walcher}. 

By the formula \eqref{5T}, we can obtain the following real BPS invariants (up to overall normalization): 
\beq
n_1^{(0,\text{real})}=12, \ 
n_3^{(0,\text{real})}=556, \ 
n_5^{(0,\text{real})}=205552, \ 
n_7^{(0,\text{real})}=121112796, .... 
\eeq
The real BPS invariants of higher degrees are listed in appendix \ref{invariants}.

\subsubsection{Small moduli limit and Self-duality}
Then we turn to study at the small moduli limit and 
confirm the self-dual property. 
As above we use the notation \eqref{A}. 
We turn back to the period integral \eqref{5period}: 
\begin{align}
\Pi&=\frac{1}{2}\int \frac{dTd\zeta}{(2\pi i)^4T\zeta}dSdX\frac{1}{[(1-X)(1-S)+z^{1/2}A(T,\zeta)SX(1-SX)]}. 
\end{align}
In the small moduli limit $z^{-1/2}\ll 1$, 
by considering the same transformations as \eqref{Tzeta}, 
we find 
\beq
\Pi=-\frac{z^{-1/2}}{2}\int \frac{dT'd\zeta'}{(2\pi i)^4T'\zeta'}dSdX\frac{1}{[SX(1-SX)+2^{-4}z^{-1/2}A'(T',\zeta')(1-X)(1-S)]}, 
\eeq
where $A'(T',\zeta')=({T'}^{1/2}+{T'}^{-1/2})(\zeta'+{\zeta'}^{-1})$. 
In the following, we drop the prime symbol $'$ for simplicity. 

Performing the coordinate transformations, 
$X\longrightarrow (X-1)/(SX)$ and $S\longrightarrow 1-S$ in turn, 
we have 
\beq
\Pi'=z^{1/2}\Pi=\frac{1}{2}\int \frac{dTd\zeta}{(2\pi i)^4T\zeta}dSdX\frac{1}{[(1-S)(1-X)-(2^8z)^{-1/2}A(T,\zeta)SX(1-SX)]}. 
\eeq

Thus, we have exactly verified the self duality of $\check{X}_5$ under 
\beq
\Pi'=z^{1/2}\Pi, \ \ x=-1/(2^{8}z). 
\eeq
$x$ is a good genuine moduli parameter in this small moduli limit 
and the shift $\Theta\longrightarrow \Theta+1/2$ noted in \cite{Kanazawa} can be naturally explained as the gauge transformation of the period, $\Pi\longrightarrow \Pi'=z^{1/2}\Pi$.

\subsection{Mirror of degree $7$}
The pfaffian Calabi-Yau $\check{X}_{7}$ is the model No.109 in \cite{data}. 

We choose the $P_2$, $P_3$ and $P_4$ for constructing the holomorphic $3$-form: 
\begin{align}
&P_2=-x_3x_4x_5+\frac{1}{\psi}x_6^2+\frac{1}{\psi^2}(x_0+x_1)x_2^3, \\
&P_3=-x_0x_1x_3x_4+\frac{1}{\psi}x_2^4+\frac{1}{\psi^2}x_5x_6, \\
&P_4=-x_0x_1x_6+\frac{1}{\psi}x_5^2+\frac{1}{\psi^2}x_2^3(x_3+x_4), \\
&P_{234}=\frac{1}{\psi}x_2^3. 
\end{align}
The hyperplanes we choose is $x_0+x_1=x_3+x_4=0$ and the defining equations of the brane are 
\beq
x_3^2x_5+\frac{1}{\psi}x_6^2=0, \ \ 
x_0^2x_3^2-\frac{1}{\psi}x_2^4-\frac{1}{\psi^2}x_5x_6=0, \ \ 
x_0^2x_6+\frac{1}{\psi}x_5^2=0. 
\eeq
By introducing new coordinates as follows: 
\beq
\frac{x_3x_4x_5}{x_6^2}=\frac{1}{\psi}W^2, \ \ 
\frac{x_0x_1x_6}{x_5^2}=\frac{1}{\psi}U, \ \ 
\frac{x_5x_6}{x_2^4}=\psi\frac{1}{S}, 
\eeq
we have the following transformation law: 
\begin{align}
&x_0=1, \ \ x_1=T, \\
&x_2=\psi^{1/4}Y^{a/2}S^{1/4}T^{1/4}U^{-1/4}W^{-1/2}, \\
&x_3=Y^{a}\zeta, \ \ x_4=Y^{a}\zeta^{-1}, \\
&x_5=\psi Y^{2a/3}T^{2/3}U^{-2/3}W^{-2/3}, \ \ 
x_6=\psi Y^{4a/3}T^{1/3}U^{-1/3}W^{-4/3}. 
\end{align}
The defining equations in the new coordinates are ($a$ is fixed to $6/7$) 
\begin{align}
&P_2=\psi Y^{16/7}T^{2/3}U^{-2/3}W^{-8/3}
[1-W^2+\psi^{-9/4}(1+T)Y^{-1}S^{3/4}T^{1/12}U^{-1/12}W^{7/6}], \\
&P_3=Y^{12/7}STU^{-1}W^{-2}S^{-1}[1+S-UW^2], \\
&P_4=\psi Y^{8/7}T^{4/3}U^{-4/3}W^{-4/3}
[1-U+\psi^{-9/4}(\zeta+\zeta^{-1})YS^{3/4}T^{-7/12}U^{7/12}W^{-1/6}]. 
\end{align}
The defining equations of the brane in the new coordinates are 
\bea
1+T=0, \ \ Y^{6/7}(\zeta+\zeta^{-1})=0, \ \ U=1, \ \ 1+S-UW^2=0, \ \ W=\pm 1. 
\ena

Now, we introduce a genuine moduli $z=\psi^{-9}$ and a new coordinate $X=W^2$. 
Then, 
\begin{align}
\Pi= &\frac{7}{(2\pi i)^6}\int\frac{P_{234}}{P_2P_3P_4}\omega_0\nonum\\
=&\frac{1}{(2\pi i)^6}\int\frac{dTd\zeta dY}{T\zeta Y}dSdUdX
\frac{1}{[1-X+z^{-1/4}(1+T)Y^{-1}S^{3/4}T^{1/12}U^{-1/12}X^{7/12}]}\nonum\\
&\times\frac{1}{[1+S-UX]}
\frac{1}{[1-U+z^{-1/4}(\zeta+\zeta^{-1})YS^{3/4}T^{-7/12}U^{7/12}X^{-1/12}]}. 
\end{align}
For the $S$-integration, we perform a residue integral at the pole $S=UX-1$, 
and the integral becomes 
\begin{align}
\Pi=&\frac{1}{(2\pi i)^5}\int\frac{dTd\zeta dY}{T\zeta Y}dUdX
\frac{1}{[1-X+z^{1/4}(1+T)Y^{-1}(UX-1)^{3/4}T^{1/12}U^{-1/12}X^{7/12}]}\nonum\\
&\times\frac{1}{[1-U+z^{1/4}(\zeta+\zeta^{-1})(UX-1)^{3/4}YT^{-7/12}U^{7/12}X^{-1/12}]}. 
\end{align}
For the $Y$-integration, we pick up the pole at 
$Y=\frac{U-1}{z^{1/4}(\zeta+\zeta^{-1})(UX-1)^{3/4}T^{-7/12}U^{7/12}X^{-1/12}}$, 
and we find 
\begin{align}
\Pi=&-\frac{1}{(2\pi i)^4}\int\frac{dTd\zeta}{T\zeta}dUdX
\frac{1}{[(1-X)(1-U)-z^{1/2}A(T,\zeta)(UX-1)^{3/2}U^{1/2}X^{1/2}]}. \label{7period}
\end{align}
Then, expanding this around the large moduli limit point ($z^{1/2}\ll 1$), we have 
\begin{align}
\Pi=&-\sum_{n=0}^{\infty}z^{n/2}\int\frac{dTd\zeta}{(2\pi i)^2T\zeta}A(T,\zeta)^n
\int \frac{dUdX}{(2\pi i)^2}
\frac{(UX-1)^{3n/2}U^{n/2}X^{n/2}}{(1-X)^{n+1}(1-U)^{n+1}}. 
\end{align}
Under the following binomial expansion: 
\beq
(UX-1)^{3n/2}=\sum_{k}^{}\frac{\Gamma(3n/2+1)}{\Gamma(k+1)\Gamma(3n/2-k+1)}{U}^{3n/2-k}{X}^{3n/2-k}(-1)^{k}, 
\eeq
and performing the analytic continuation, we obtain the factorized integral 
\begin{align}
\Pi=&-\int\frac{ds}{2\pi i}\frac{\pi\cos(\pi s)}{\sin(\pi s)}z^{s/2}\int\frac{dTd\zeta}{(2\pi i)^2T\zeta}A(T,\zeta)^{s}
\sum_{k}^{}\frac{\Gamma(3s/2+1)(-1)^k}{\Gamma(k+1)\Gamma(3s/2-k+1)}\nonum\\
&\times
\int\frac{dU}{2\pi i}U^{2s-k}(1-U)^{-s-1}
\int\frac{dX}{2\pi i}X^{2s-k}(1-X)^{-s-1}. 
\end{align}
The $T$- and $\zeta$-part are exactly the same as in the other models, \eqref{T} and \eqref{zeta}. 
The $U$- and $X$-part can be evaluated by using \eqref{gamma}.

Collecting all results, we find simple poles at $s=2n$, 
and the fundamental period is 
\begin{align}
\Pi_0
&=\sum_{n=0}^{\infty}z^n\frac{\Gamma(2n+1)^2}{\Gamma(n+1)^4}\sum_{k=0}^{2n}(-1)^k
\frac{\Gamma(3n+1)\Gamma(4n-k+1)^2}{\Gamma(3n-k+1)\Gamma(2n-k+1)^2\Gamma(2n+1)^2\Gamma(k+1)}\nonum\\
&=\sum_{n=0}^{\infty}z^n
\begin{pmatrix} 2n \\ n \end{pmatrix}^2
\sum_{k=0}^{2n}(-1)^k
\begin{pmatrix} 3n \\ k \end{pmatrix}
\begin{pmatrix} 4n-k \\ 2n \end{pmatrix}^2
. \label{7funda}
\end{align}
We can show that the above formula is the same one obtained in \cite{Kanazawa} by the use of hypergeometric identities.

For the domainwall tension (precisely $\mathcal{W}_+$), 
by replacing the contour integral with respect to $X$ with the line integral, 
we find simple poles at $s=2n+1$ and double poles at $s=2n$. 
As a result, we obtain $\mathcal{W}_+=\frac{1}{4\pi i}\Pi_1+\frac{1}{4}\Pi_0+\frac{1}{2}\tau$, 
where $\Pi_0$ is the fundamental period given in \eqref{7funda}, 
$\Pi_1$ is the logarithmic period given by 
\begin{align}
\Pi_1=&\Pi_0\log z+\sum_{n=0}^{\infty}z^n
\begin{pmatrix} 2n \\ n \end{pmatrix}^2
\sum_{k=0}^{2n}(-1)^k
\begin{pmatrix} 3n \\ k \end{pmatrix}
\begin{pmatrix} 4n-k \\ 2n \end{pmatrix}^2\nonum\\
&\times[-4\Psi(n+1)+3\Psi(3n+1)-3\Psi(3n-k+1)+8\Psi(4n-k+1)-4\Psi(2n-k+1)], 
\end{align}
and $\tau$ is given by 
\beq
\tau=\sum_{n=0}^{\infty}z^{n+1/2}
\frac{\Gamma(2n+2)^2}{\Gamma(n+3/2)^4}
\sum_{k=0}^{2n+1}
\frac{(-1)^k\Gamma(3n+5/2)\Gamma(4n-k+3)^2}{\Gamma(3n-k+5/2)\Gamma(k+1)\Gamma(2n-k+2)^2\Gamma(2n+2)^2}. 
\eeq

By the use of the mirror map, 
we can obtain the prediction for the real BPS invariants (up to overall normalization), as follows: 
\beq
n_1^{(0,\text{real})}=10, \ \ 
n_3^{(0,\text{real})}=204, \ \ 
n_5^{(0,\text{real})}=43790, \ \ 
n_7^{(0,\text{real})}=14034754, \ \ 
n_9^{(0,\text{real})}=5377152402, \ .... 
\eeq
The real BPS invariants of higher degrees are listed in appendix \ref{invariants}.

\subsection{Mirror of degree $10$}
For the large moduli, the pfaffian Calabi-Yau $\check{X}_{10}$ is the model No.263, and for the small moduli, this is the model No.271, as listed in \cite{data}. 

We choose $P_0$, $P_3$ and $P_4$ for constructing the holomorphic $3$-form: 
\begin{align}
&P_0=-x_2x_3x_4+\frac{1}{\psi}x_5^3+\frac{1}{\psi^2}(x_0+x_1)x_6, \\
&P_3=-x_0x_1x_5+\frac{1}{\psi}x_4^3+\frac{1}{\psi^2}(x_2+x_3)x_6, \\
&P_4=-x_0x_1x_2x_3+\frac{1}{\psi}x_6^2+\frac{1}{\psi^2}x_4^2x_5^2, \\
&P_{034}=\frac{1}{\psi}x_6. 
\end{align}
The hyperplanes are $x_0+x_1=x_2+x_3=0$ and the defining equations of the brane are 
\beq
x_2^2x_4+\frac{1}{\psi}x_5^3=0, \ \ 
x_0^2x_5+\frac{1}{\psi}x_4^3=0, \ \ 
x_0^2x_2^2-\frac{1}{\psi}x_6^2-\frac{1}{\psi^2}x_4^2x_5^2=0. 
\eeq
By introducing new coordinates for each invariant monomial as follows: 
\beq
\frac{x_2x_3x_4}{x_5^3}=\frac{1}{\psi}\frac{1}{U}, \ \ 
\frac{x_0x_1x_5}{x_4^3}=\frac{1}{\psi}\frac{1}{W^2}, \ \ 
\frac{x_4^2x_5^2}{x_6^2}=\psi S, 
\eeq
we obtain the following transformation law: 
\begin{align}
&x_0=1, \ \ x_1=T, \ \ 
x_2=Y^{a}\zeta, \ \ x_3=Y^{a}\zeta^{-1}, \\
&x_4=\psi^{1/2} Y^{a/4}T^{3/8}U^{1/8}W^{3/4}, \\
&x_5=\psi^{1/2} Y^{3a/4}T^{1/8}U^{3/8}W^{1/4}, \\ 
&x_6=\psi^{1/2} Y^{a}S^{-1/2}T^{1/2}U^{1/2}W^{}. 
\end{align}
The defining equations in the above new coordinates are as follows ($a$ is fixed to $4/5$): 
\begin{align}
P_0&=\psi^{1/2} Y^{9/5}T^{3/8}U^{1/8}W^{3/4}
[U-1+\psi^{-2}(1+T)Y^{-1}S^{-1/2}T^{1/8}U^{3/8}W^{1/4}], \\
P_3&=\psi^{1/2} Y^{3/5}T^{9/8}U^{3/8}W^{1/4}[W^2-1+\psi^{-2}(\zeta+\zeta^{-1}) YS^{-1/2}T^{-5/8}U^{1/8}W^{3/4}], \\
P_4&=TY^{8/5}S^{-1}[S(UW^2-1)+UW^2]. 
\end{align}
The defining equations of the brane in the new coordinates are 
\beq
1+T=0, \ \ Y^{4/5}(\zeta+\zeta^{-1})=0, \ \ U=1, \ \ S(UW^2-1)+UW^2=0, \ \ W=\pm 1. 
\eeq

Now we introduce a genuine moduli $z=\psi^{-8}$ and a new coordinate $X=W^2$. 
Then, the period integral is 
\begin{align}
\Pi= &\frac{10}{(2\pi i)^6}\int\frac{P_{034}}{P_0P_3P_4}\omega_0\nonum\\
=&\frac{1}{(2\pi i)^6}\int\frac{dSdTdUdXd\zeta dY}{ST\zeta Y}\frac{1}{[U-1+z^{1/4}(1+T) Y^{-1}S^{-1/2}T^{1/8}U^{3/8}X^{1/8}]}\nonum\\
&\times\frac{1}{[X-1+z^{1/4} (\zeta+\zeta^{-1})YS^{-1/2}T^{-5/8}U^{1/8}X^{3/8}]}\frac{1}{[S(UX-1)+UX]}. 
\end{align}
For the $S$-integration, we pick up the pole at $S=\frac{UX}{1-UX}$, 
and the integral becomes 
\begin{align}
\Pi=&\int\frac{dTd\zeta dYdUdX}{(2\pi i)^5T\zeta YUX}\frac{1}{[1-U-z^{1/4}(1+T)(1-UX)^{1/2}Y^{-1}T^{1/8}U^{-1/8}X^{-3/8}]}\nonum\\
&\times\frac{1}{[X-1+z^{1/4}(\zeta+\zeta^{-1})(1-UX)^{1/2}YT^{-5/8}U^{-3/8}X^{-1/8}]}. 
\end{align}
For the $Y$-integration, by picking up the pole at 
$Y=\frac{1-X}{z^{1/4}(\zeta+\zeta^{-1})(1-UX)^{1/2}T^{-5/8}U^{-3/8}X^{-1/8}}$, 
we obtain 
\begin{align}
\Pi=\int\frac{dTd\zeta dUdX}{(2\pi i)^4T\zeta UX}\frac{1}{[(1-U)(1-X)-z^{1/2}A(T,\zeta)(1-UX)U^{-1/2}X^{-1/2}]}. \label{10period}
\end{align}

\subsubsection{Large moduli limit}
This is the model No.263 in \cite{data}. 
Expanding \eqref{10period} by $z^{1/2}\ll 1$, the period integral becomes 
\begin{align}
\Pi=\sum_{n=0}^{\infty}z^{n/2}\int\frac{dTd\zeta}{(2\pi i)^2T\zeta}A(T,\zeta)^n\int\frac{dUdX}{(2\pi i)^2UX}\frac{(1-UX)^nU^{-n/2}X^{-n/2}}{(1-U)^{n+1}(1-X)^{n+1}}. 
\end{align}
Then, using the following binomial expansion formula: 
\beq
(1-UX)^{n}=\sum_{k}^{}\frac{\Gamma(n+1)}{\Gamma(n-k+1)\Gamma(k+1)}(-1)^kU^kX^k, 
\eeq
and performing the analytic continuation, 
we obtain the factorized integral 
\begin{align}
\Pi=&\int\frac{ds}{2\pi i}\frac{\pi\cos(\pi s)}{\sin(\pi s)}z^{s/2}
\int\frac{dTd\zeta}{(2\pi i)^2T\zeta}A(T,\zeta)^{s}
\sum_{k}^{}\frac{\Gamma(n+1)(-1)^k}{\Gamma(n-k+1)\Gamma(k+1)}\nonum\\
&\times
\int\frac{dU}{2\pi i}U^{-s/2+k-1}(1-U)^{-s-1}
\int\frac{dX}{2\pi i}X^{-s/2+k-1}(1-X)^{-s-1}. 
\end{align}
The $T$- and $\zeta$-integral are the same as in the other models, \eqref{T} and \eqref{zeta}. 
The $U$- and $X$-integral can be evaluated by using \eqref{gamma}. 

By collecting all integrals and performing the $s$-integration, we find poles at $s=2n$ and obtain 
\begin{align}
\Pi_0
=&\sum_{n=0}^{\infty}z^n
\begin{pmatrix} 2n \\ n \end{pmatrix}^2\sum_{k=0}^{n}\frac{(-1)^k\Gamma(3n-k+1)^2}{\Gamma(2n-k+1)\Gamma(n-k+1)^2\Gamma(2n+1)\Gamma(k+1)}\nonum\\
=&\sum_{n=0}^{\infty}z^n
\begin{pmatrix} 2n \\ n \end{pmatrix}^2\sum_{k=0}^{n}(-1)^k
\begin{pmatrix} 2n \\ k \end{pmatrix}
\begin{pmatrix} 3n-k \\ 2n \end{pmatrix}^2. \label{10funda}
\end{align}
We can show that the above formula is the same one obtained in \cite{Kanazawa} by the use of hypergeometric identities.

Then, for the domainwall tension (precisely $\mathcal{W}_+$), 
by replacing the contour integral with respect to $X$ with the line integral, 
we find simple poles at $s=2n+1$ and double poles at $s=2n$. 
As a result, we obtain $\mathcal{W}_+=\frac{1}{4\pi i}\Pi_1+\frac{1}{4}\Pi_0+\frac{1}{2}\tau$, 
where $\Pi_0$ is the fundamental period given in \eqref{10funda}, 
$\Pi_1$ is the logarithmic period given by 
\begin{align}
\Pi_1=&\Pi_0\log z+
\sum_{n=0}^{\infty}z^n
\begin{pmatrix} 2n \\ n \end{pmatrix}^2
\sum_{k=0}^{n}(-1)^k
\begin{pmatrix} 2n \\ k \end{pmatrix}
\begin{pmatrix} 3n-k \\ 2n \end{pmatrix}^2
[2\Psi(2n+1)-4\Psi(n+1)\nonum\\
&-2\Psi(2n-k+1)+6\Psi(3n-k+1)-2\Psi(n-k+1)], 
\end{align}
and $\tau$ is given by 
\begin{align}
\tau=\sum_{n=0}^{\infty}z^{n+1/2}
\frac{\Gamma(2n+2)^2}{\Gamma(n+3/2)^4}
\sum_{k=0}^{2n+1}(-1)^k\frac{\Gamma(3n-k+5/2)^2}{\Gamma(2n-k+2)\Gamma(2n+2)\Gamma(n-k+3/2)^2\Gamma(k+1)}. 
\end{align}

Under the A-model interpretation by the mirror map, we find the following real BPS invariants (up to overall normalization): 
\beq
n_{1}^{(0,\text{real})}=8, \ 
n_{3}^{(0,\text{real})}=72, \ 
n_{5}^{(0,\text{real})}=7840, \ 
n_{7}^{(0,\text{real})}=1275496, \ 
n_{9}^{(0,\text{real})}=243115152, \ .... 
\eeq
The real BPS invariants of higher degrees are listed in appendix \ref{invariants}.

\subsubsection{Small moduli limit}
Then we consider the small moduli limit, $z^{-1/2}\ll 1$. 
This is the model No.271 in \cite{data}. 
We restart by the formula \eqref{10period}: 
\begin{align}
\Pi=\int\frac{dTd\zeta dUdX}{(2\pi i)^4T\zeta UX}\frac{1}{[(1-U)(1-X)-z^{1/2}A(T,\zeta)(1-UX)U^{-1/2}X^{-1/2}]}. 
\end{align}
By performing the same transformations with respect to $T$ and $\zeta$ as those in the small moduli limit of $\check{X}_{13}$ and $\check{X}_5$, and by dropping the prime symbol $'$, 
we have 
\begin{align}
\Pi=-z^{-1/2}\int\frac{dTd\zeta dUdX}{(2\pi i)^4T\zeta U^{1/2}X^{1/2}}\frac{1}{[(1-UX)-2^{-4}z^{-1/2}A(T,\zeta)(1-U)(1-X)U^{1/2}X^{1/2}]}. 
\end{align}
The shift $\Theta\longrightarrow \Theta+1/2$ noted in \cite{Kanazawa} can be naturally explained as the gauge transformation of the period, $\Pi\longrightarrow \Pi'=z^{1/2}\Pi$. 

Performing the coordinate transformations, 
$U\longrightarrow U/X$ and $X\longrightarrow 1/(1-X)$ in turn, 
the integral becomes 
\beq
\Pi'=\int\frac{dTd\zeta dUdX}{(2\pi i)^4T\zeta {U}^{1/2}}\frac{1}{[(1-U)(1-X)+2^{-4}z^{-1/2}A(T,\zeta)((1-U)+UX)X{U}^{1/2}]}. \label{10smallperiod}
\eeq
By expanding \eqref{10smallperiod} around the small moduli limit point, 
we obtain 
\beq
\Pi'=\sum_{n=0}^{\infty}(-2^{-4}z^{-1/2})^{n}\int\frac{dTd\zeta}{(2\pi i)^2T\zeta}A(T,\zeta)^{n}\int\frac{dUdX}{(2\pi i)^2}\frac{((1-U)+UX)^nX^nU^{n/2-1/2}}{(1-U)^{n+1}(1-X)^{n+1}}. 
\eeq
Under the following binomial expansion: 
\bea
((1-U)+UX)^{n}=\sum_{k}\frac{\Gamma(n+1)}{\Gamma(k+1)\Gamma(n-k+1)}
(1-U)^{k}U^{n-k}X^{n-k}, 
\ena
and performing the analytic continuation, we obtain the factorized integral 
\begin{align}
\Pi'= &\int\frac{ds}{2\pi i}\frac{\pi\cos(\pi s)}{\sin(\pi s)}
(-2^{-4}z^{-1/2})^{s}\int\frac{dTd\zeta}{(2\pi i)^2T\zeta}A(T,\zeta)^{s}
\sum_{k}\frac{\Gamma(s+1)}{\Gamma(k+1)\Gamma(s-k+1)}\nonum\\
&\times\int \frac{dU}{2\pi i}U^{3s/2-k-1/2}(1-U)^{-s+k-1}\int \frac{dX}{2\pi i}X^{2s-k}(1-X)^{-s-1}. 
\end{align}

First we consider the fundamental period. 
The $T$- and $\zeta$-integral are exactly the same as in the other models and given by \eqref{T} and \eqref{zeta}, respectively. 
The $U$- and $W$-part can be evaluated by using \eqref{gamma}. 

Now we introduce the genuine moduli as $x=-2^{-12}z^{-1}$. 
Then the fundamental period becomes 
\bea
\Pi_0'=\sum_{n=0}^{\infty}{x}^{n}2^{4n}
\begin{pmatrix}
2n \\ n
\end{pmatrix}^2
\sum_{k=0}^{2n}(-1)^k\frac{\Gamma(3n-k+1/2)\Gamma(4n-k+1)}{\Gamma(n+1/2)\Gamma(2n-k+1)^3\Gamma(k+1)}. \label{10smallfunda}
\ena

For the domainwall tension (precisely $\mathcal{W}_+'$), 
by replacing the contour integral with respect to $X$ with the line integral, 
we find simple poles at $s=2n+1$ and double poles at $s=2n$. 
As a result, we obtain $\mathcal{W}_+'=\frac{1}{4\pi i}\Pi_1'+\frac{1}{4}\Pi_0'+\frac{1}{2}\tau'$, 
where $\Pi_0'$ is the fundamental period given in \eqref{10smallfunda}, 
$\Pi_1'$ is the logarithmic period given by 
\begin{align}
\Pi_1'=&\Pi_0' \log z+\sum_{n=0}^{\infty}x^{n}2^{4n}
\begin{pmatrix}
2n \\ n
\end{pmatrix}^2
\sum_{k=0}^{2n}(-1)^k\frac{\Gamma(3n-k+1/2)\Gamma(4n-k+1)}{\Gamma(n+1/2)\Gamma(2n-k+1)^3\Gamma(k+1)}\nonum\\
&\times [4\Psi(2n+1)-4\Psi(n+1)+3\Psi(3n-k+1/2)+4\Psi(4n-k+1)\nonum\\
&-\Psi(n+1/2)-6\Psi(2n-k+1)+2\Psi(1)-2\Psi(1/2)], 
\end{align}
and $\tau'$ is given by 
\beq
\tau'=\sum_{n=0}^{\infty}x^{n+1/2}2^{4n+2}\frac{\Gamma(2n+2)^2}{\Gamma(n+3/2)^4}\sum_{k=0}^{2n+1}(-1)^k
\frac{\Gamma(3n-k+2)\Gamma(4n-k+3)}{\Gamma(n+1)\Gamma(2n-k+2)^3\Gamma(k+1)}. 
\eeq

Under the mirror map, we can obtain the following real BPS invariants (up to overall normalization): 
\beq
n_1^{(0,\text{real})}=16, \ 
n_3^{(0,\text{real})}=8048, \ 
n_5^{(0,\text{real})}=12744560, \ 
n_7^{(0,\text{real})}=34858414832, \ .... 
\eeq
The real BPS invariants of higher degrees are listed in appendix \ref{invariants}.

\subsection{Inhomogeneous term}\label{inhomo}
Here we present the inhomogeneous terms of Picard-Fuchs equations for all models $\check{X}_i$. 
For pfaffians treated in this paper, as explained in section \ref{hol3_period}, the inhomogeneous term is in general written in the following form: 
\beq
\mathcal{L}_{\text{PF}}\mathcal{T}=c_1z^{1/2}+c_2z^{3/2}+c_3z^{5/2}, \nonum
\eeq
where $c_i$ $(i=1,2,3)$ are certain constants. 
It is very interesting that there is such a difference from the well-known hypersurface/complete intersection cases. 
We list the concrete form of the inhomogeneous terms for the pfaffians in Table \ref{inhomogeneous}. 
\begin{table}
\begin{center}
\begin{tabular}{l|cl} 
\toprule
$\check{X}_i$ & limit & inhomogeneous term \\ 
\midrule
$\check{X}_{13}$ & large & $\frac{1183}{16} z^{1/2}+5278 z^{3/2}+4272 z^{5/2}$ \\ [-3pt] 
 & small & $ 4x^{1/2}-2031616x^{3/2}+28789702656x^{5/2}$ \\ [-3pt] 
\midrule
$\check{X}_{5}$ & & $\frac{93}{4} z^{1/2}+7440 z^{3/2}+49600 z^{5/2}$ \\ [-3pt] 
\midrule
$\check{X}_{7}$ & & $\frac{245}{8}z^{1/2}+4326z^{3/2}+10638z^{5/2}$ \\ [-3pt] 
\midrule
$\check{X}_{10}$ & large & $\frac{25}{2} z^{1/2}+1240 z^{3/2}+2048 z^{5/2}$ \\ [-3pt] 
 & small & $x^{1/2}+17408 x^{3/2}+7536640 x^{5/2}$ \\ 
\bottomrule
\end{tabular}
\end{center}
\caption{Inhomogeneous term}\label{inhomogeneous}
\end{table}
Here we use the normalization \eqref{normalization}. 
These inhomogeneous terms are evaluated simply by acting the Picard-Fuchs operators on the domainwall tensions, which have been obtained by the direct integration method in this section. 
It seems tedious to carry out the computation of the inhomogeneous term for the pfaffians via the method in \cite{MW}.

\section{One-loop Consideration}\label{one-loop}
In this section, we discuss the one-loop amplitudes and the one-loop real BPS invariants. 
The holomorphic anomaly equation relates the tree level amplitudes to the amplitudes of the higher worldsheet topologies \cite{BCOV1,BCOV2}. 
The extension to the open string sector of the holomorphic anomaly equation, in particular for the compact Calabi-Yau case, was proposed in \cite{Wa2,Wa3}. 
Under certain additional conditions (discreteness of open string moduli, tadpole cancellation, e.t.c. \cite{Wa3}), 
the amplitudes for higher worldsheet topology are then constrained recursively by the extended holomorphic anomaly equation \cite{Wa2}. 
By using the formula for one loop amplitudes given in these works, 
we will present further possible enumerative predictions as well as additional consistency checks. 

\subsection{One-loop amplitude}
In the previous section, the genus $0$ real BPS invariants of several pfaffians  have been predicted up to an overall numerical normalization. 
The direct evaluation of these invariants by a localization computation in the topological A-model helps us to fix the normalization ambiguities. 
Now, instead of such arguments, 
we try to determine the overall normalization factor of the mirror relation 
by concerning the one-loop amplitudes. 
We must fix the holomorphic ambiguity which cannot be determined by the holomorphic anomaly equation itself.  

In \cite{Wa3}, it was showed that the following combination of $1$-loop amplitudes has the BPS expansions: 
\beq
\mathcal{A}^{\text{hol}}+\mathcal{K}^{\text{hol}}=\sum_{d:\text{even}>0, \ k:\text{odd}>0}\frac{2n_d^{(1,\text{real})}}{k}q^{dk/2}, 
\eeq
where $\mathcal{A}^{\text{hol}}$ is the holomorphic limit of the annulus amplitude and $\mathcal{K}^{\text{hol}}$ is that of the Klein bottle amplitude. 
$n_d^{(1,\text{real})}$ are genus $1$ real BPS invariants and all of them are expected to be integers. 

For all the well-known hypersurface/complete intersection models, by fixing holomorphic ambiguities, 
$\mathcal{A}^{\text{hol}}$ and $\mathcal{K}^{\text{hol}}$ are given by 
\begin{align}
&\partial_z \mathcal{A}^{\text{hol}}=-\frac{1}{2}(\Delta_{zz}^{\text{hol}})^2C^{-1}_{zzz}, \nonum\\ 
&\mathcal{K}^{\text{hol}}=\frac{1}{2}\log\left[\frac{q}{z}\frac{dz}{dq}z^{-1}(\text{diss}1)^{-1/4}\right], \label{AZ}
\end{align}
where 
\begin{align}
\Delta_{zz}^{\text{hol}}
=\left(\partial_z-\partial_z\log\frac{dt(z)}{dz}-\partial_z\log\Pi_0\right)(\partial_z-\partial_z\log\Pi_0)\mathcal{T}(z), 
\end{align}
$C_{zzz}$ is the quantum Yukawa coupling \cite{Wa3,KW1,KS1}, 
and $(\text{diss}1)$ is the discriminant. 

In the pfaffian cases, the quantum Yukawa coupling takes the following form: 
\beq
C_{zzz}=\text{Const.}\frac{(\text{diss}2)}{z^3(\text{diss}1)}. \label{quantum_yukawa}
\eeq
Here, $(\text{diss}1)$ and ($\text{diss}2$) are listed in Table \ref{moduli} and can be obtained from the Picard-Fuchs operators easily. 
A numerical constants can be fixed by the condition that $C_{zzz}$ has the form 
\beq
C_{zzz}\sim\text{degree}+\mathcal{O}(z)+...
\eeq
around $z\sim 0$. 
Since the pfaffians, unlike the well-known hypersurfaces/complete intersections, have non-trivial numerator of the quantum Yukawa coupling, 
it is possible that additional ambiguities and modifications appear in \eqref{AZ}. 
Indeed, for some of the models, the use of \eqref{AZ} does not give integral invariants under the A-model interpretation. 

In this paper, we consider the following ansatz: 
\begin{align}
&\tau \longrightarrow a\tau \ \ 
(\Longrightarrow \ \mathcal{A}\longrightarrow a\mathcal{A}), \\
&\mathcal{K}^{\text{hol}}=\frac{1}{2}\log\left[\frac{q}{z}\frac{dz}{dq}z^{-1}\left(\frac{(\text{diss}2)}{(\text{diss1})}\right)^{b}\right], \label{ansatz1}
\end{align}
where $a$ and $b$ are rational numbers. 
Or, instead of \eqref{ansatz1}, we consider 
\beq
\mathcal{K}^{\text{hol}}=\frac{1}{2}\log\left[\frac{q}{z}\frac{dz}{dq}z^{-1}\frac{(\text{diss}2)^{c}}{(\text{diss}1)^{1/4}}\right], 
\eeq
where $c$ is a rational number. 
Needless to say, we could have other possibilities of holomorphic ambiguities which may change the form of one-loop amplitudes. 
We explore possibilities of $(a,b)$ or $(a,c)$ which exhibit integral expansions at both the tree and the one-loop levels. 
It is important that we have a possibility $a=0$, namely, 
all of disk invariants and annulus invariants are zero. 
Then, one-loop BPS invariants just come from the contributions of Klein bottle invariants. 

We have several additional consistency conditions. 
It was showed in \cite{PaSoWa}
\beq
n_{d}^{(0,\text{real})}=\frac{1}{2}n_{d}^{(0,\text{disk})}, \label{C1}
\eeq
namely, the number of real BPS invariants at the tree level is the half of the number of holomorphic discs. 
Moreover, as noted in \cite{Wa3,KW1}, we have 
\begin{align}
n_{d}^{(\hat{g},\text{real})}\equiv n_{d}^{(\hat{g})} \ \ \ (mod \ 2), \label{C2}\\ 
n_{d}^{(\hat{g})}=0\Longrightarrow n_{d}^{(\hat{g},\text{real})}=0. \label{C3}
\end{align}
In the following, we will try to fix overall normalizations 
by using the above ansatz and consistency conditions.

\subsection{One-loop BPS invariants}
\subsubsection{Degree $13$}
\begin{itemize}
\item \textbf{Large moduli} 
\end{itemize}
In this case, $(a,b)=(1,1/4)$ (equivalently, $c=1/4$) seems to be the unique possibility. 
The one-loop invariants are as follows: 
\begin{center}
\begin{tabular}{l|ccccccc} \toprule
$(a,b)$ & $d=2$ & $d=4$ & $d=6$ & $d=8$ & $d=10$ & $d=12$ & ... \\ 
\midrule
$(1,1/4)$ & $0$ & $120$ & $36576$ & $8017614$ & $1653968592$ & $336816482863$ & ... \\
\bottomrule
\end{tabular}
\end{center}
This is natural in the sense 
that the one-loop amplitude in the pfaffian case can be obtained 
as that in the well-known one-loop formula for hypersurface/complete intersection cases \eqref{AZ}, 
by simply replacing $1/(\text{diss}1)$ with $(\text{diss}2)/(\text{diss}1)$, 
just like the quantum Yukawa coupling. 
So one might expect that this is the general property for all the pfaffians we treat. 
But it will turn out that this simple modification does not work for some of the other models ($\check{X}_{7}$ and $\check{X}_{10}$) and we must consider other possibilities. 

\begin{itemize}
\item \textbf{Small moduli} 
\end{itemize}
In this case, we have so many possibilities which exhibit integral invariants and satisfying the consistency conditions \eqref{C1}, \eqref{C2} and \eqref{C3}. 
So it is hard to fix normalization completely by this consideration. 
We list several possibilities below: 
\begin{center}
\begin{tabular}{l|ccccc} \toprule
$(a,b)$ & $d=2$ & $d=4$ & $d=6$ & $d=8$ & ... \\ 
\midrule
$(13,1/4)$ & $-39936$ & $1699435056$ & $-94316793219584$ & $5006546796335364656$ & ... \\[-10pt] 
$(0,1/4)$ & $3328$ & $-125267408$ & $9017682342656$ & $-362000262411814352$ & ... \\[-10pt] 
... & & & & & \\ 
\midrule
$(a,c)$ & $d=2$ & $d=4$ & $d=6$ & $d=8$ & ... \\ 
\midrule
$(0,0)$ & $0$ & $37671472$ & $1255342506496$ & $65370203450538544$ & ... \\[-10pt] 
$(13,1/2)$ & $-36608$ & $1536496176$ & $-86554453383424$ & $4579176330473011760$& ... \\[-10pt] 
... & & & & & \\ 
\bottomrule
\end{tabular}
\end{center}
$b=1/4$ seems to be natural since this Calabi-Yau is realized as the A-model side of the small moduli limit of the degree $13$ model $\check{X}_{13}$, and one-loop amplitudes of these two limits are expected to be connected each other by a simple relation, namely, $z\longleftrightarrow x$ and $\Pi_0\longleftrightarrow \Pi_0'$, as discussed for the closed one-loop amplitude of the case of the degree $14$ in \cite{HoKo}. 
If $b$ is fixed to this value, $a=13$, $13/2$, $13/4$, $0$, ..., show integral invariants at the tree and the one-loop levels. 
The possibility $a=0$ is the situation where all of disk invariants and annulus invariants are $0$. 
We have many other possibilities, for example, $(a,c)=(0,0)$, $(13,1/2)$, $(13/2,1/2)$, $(13,0)$, $(26,1/2)$, ..., and it is hard to decide a overall factor completely by this consideration.

\subsubsection{Degree $5$}
In this case, there are also many possibilities which give integral invariants as listed below. 
\begin{center}
\begin{tabular}{l|ccccccc} \toprule
$(a,b)$ & $d=2$ & $d=4$ & $d=6$ & $d=8$ & $d=10$ & $d=12$ & ... \\ 
\midrule
$(5,1/4)$ & $-36$ & $417$ & $4903388$ & $7140650773$ & $9221277658128$ & $11388045762256931$ & ... \\[-10pt] 
$(5,1/2)$ & $86$ & $66281$ & $59151354$ & $58611169973$ & $61567266679256$ & $67013112045986363$ & ... \\[-10pt] 
... & & & & & & \\ 
\midrule
$(a,c)$ & $d=2$ & $d=4$ & $d=6$ & $d=8$ & $d=10$ & $d=12$ & ... \\ 
\midrule
$(0,0)$ & $10$ & $8793$ & $10934230$ & $12480840693$ & $14252296596968$ & $16421644185053963$ & ... \\[-10pt] 
$(0,1/2)$ & $8$ & $9681$ & $10868576$ & $12508692053$ & $14269232220288$ & $16435001441040179$ & ... \\[-10pt] 
... & & & & & & \\ 
\bottomrule
\end{tabular}
\end{center}
$(a,b)=(5,1/4)$ is natural since $a$ is the degree of this model and $b=1/4$ is the same as the natural one for $\check{X}_{13}$. 
We have many other possibilities as listed above. 
By the consistency condition \eqref{C2} (mod $2$ equivalence with the number of genus $1$ rational curves), 
we can exclude $(a,b)=(0,1/4)$, ..., $(a,c)=(5,0)$, $(5,1/2)$, ..., although they show integral invariants.

\subsubsection{Degree $7$}
In this case, 
we can find several possibilities which exhibit integral invariants as shown below: 
\begin{center}
\begin{tabular}{l|ccccccc} \toprule
$(a,c)$ & $d=2$ & $d=4$ & $d=6$ & $d=8$ & $d=10$ & $d=12$ & ... \\ 
\midrule
$(1,1/12)$ & $2$ & $1955$ & $1464392$ & $937055117$ & $588109462058$ & $369491842053326$ & ... \\[-10pt] 
$(3,3/4)$ & $-6$ & $1587$ & $1210520$ & $821812845$ & $529228925298$ & $337701027124910$ & ... \\[-10pt] 
... & & & & & & & \\ 
\bottomrule
\end{tabular}
\end{center}
We have several possibilities $(a,c)=(1,1/12)$, $(3,3/4)$ and 
unfortunately 
we cannot choose $b=1/4$ in this case. 
By the consistency condition \eqref{C2}, 
we can exclude $(a.b)=(1,1/12)$, $(3,3/4)$, ..., $(a,c)=(0,0)$, $(2,1/3)$, $(4,4/3)$, ..., although they show integral invariants.

\subsubsection{Degree $10$}
\begin{itemize}
\item \textbf{Large moduli} 
\end{itemize}
In this case, 
we can find several possibilities which exhibit integral invariants as shown below: 
\begin{center}
\begin{tabular}{l|ccccccc} \toprule
$(a,b)$ & $d=2$ & $d=4$ & $d=6$ & $d=8$ & $d=10$ & $d=12$ & ... \\ 
\midrule
$(2,1/2)$ & $33$ & $5109$ & $1203267$ & $317827365$ & $88973783316$ & $25850175760831$ & ... \\[-10pt] 
... & & & & & & & \\ 
\midrule
$(a,c)$ & $d=2$ & $d=4$ & $d=6$ & $d=8$ & $d=10$ & $d=12$ & ... \\ 
\midrule
$(0,0)$ & $1$ & $373$ & $161763$ & $53056293$ & $16831641236$ & $5311521613375$ & ... \\[-10pt] 
$(2,1/2)$ & $-1$ & $365$ & $148205$ & $50166405$ & $16089559084$ & $5111631748583$ & ... \\[-10pt] 
... & & & & & & \\ 
\bottomrule
\end{tabular}
\end{center}
We have possibilities $(a,b)=(2,1/2)$, $(4,2)$, ..., 
and $b=1/4$ is not suitable as is the case with $\check{X}_7$. 
There are many other possibilities, for examples, $(a,c)=(0,0)$, $(2,1/2)$, $(4,2)$, $(6,9/2)$, $(8,8)$, ..., and so on. 
We cannot narrow the list any further by the consistency checks. 

\begin{itemize}
\item \textbf{Small moduli} 
\end{itemize}
In this case, 
we can find so many possibilities which exhibit integral invariants as shown below: 
\begin{center}
\begin{tabular}{l|ccccccc} \toprule
$(a,b)$ & $d=2$ & $d=4$ & $d=6$ & $d=8$ & $d=10$ & ... \\
\midrule
$(5/2,1/4)$ & $30$ & $312240$ & $1567795510$ & $8153551664688$ & $44180874975115674$ & ... \\[-10pt] 
$(10,1/4)$ & $-720$ & $-2183760$ & $-5860628240$ & $-19078691535312$ & $-76992035691620976$ & ... \\[-10pt] 
$(10,1/2)$ & $-256$ & $-584272$ & $475607296$ & $9807715812912$ & $66101470258944256$ & ... \\[-10pt] 
... & & & & & \\ 
\midrule
$(a,c)$ & $d=2$ & $d=4$ & $d=6$ & $d=8$ & $d=10$ & ... \\ 
\midrule
$(0,0)$ & $160$ & $406960$ & $1911650080$ & $9748479198768$ & $51721089281251808$ & ... \\[-10pt] 
$(5,1/2)$ & $-200$ & $-115280$ & $233484440$ & $2927658370608$ & $20484272580081320$ & ... \\[-10pt] 
... & & & & & & \\ 
\bottomrule
\end{tabular}
\end{center}
In addition, there are so many other possibilities, for example, 
$(a,b)=(5,1/4)$, $(0,1/4)$, $(5,1/2)$, ..., 
$(a,c)=(0,1/2)$, $(5/2,1/2)$, $(0,3/4)$, $(0,1/8)$, $(5,0)$, $(5/2,0)$, .... 
Hence we cannot fix a normalization completely by this consideration. 
\\

For some of the models, we have many possibilities which give integral invariants at both the tree and the one-loop orders, and satisfy the consistency conditions. 
We also have many other possibilities which exhibit integral invariants and do not satisfy the consistent conditions. 
These results are very interesting and it is tempting to conjecture that even in such cases we count the number of certain unknown objects in Calabi-Yau manifolds. 
For fixing ambiguities of the overall normalization completely, 
it is necessary to consider the direct computations of disk invariants in the A-model side.

\section{Conclusion and Discussions}\label{conclusion}
In this paper, we investigated the open mirror symmetry for certain compact non-complete intersection Calabi-Yaus, called pfaffian Calabi-Yau $3$-folds. 
We computed the B-type background D-brane superpotential, which contributes to the space-time superpotential, via the direct integration method provided recently. 
We choose the holomorphic curve with two discrete moduli as the B-brane. 
By the use of the open mirror conjecture for compact Calabi-Yau manifolds, 
we have extracted the BPS invariants (disk invariants) up to an overall numerical normalization, and under certain choices of overall normalizations and holomorphic ambiguities, 
those invariants certainly show the integrality property.

We have discussed both the large moduli limit and the small moduli limit. 
In the degree $13$ and the degree $10$ cases, these two limits correspond to the large volume limits of two different Calabi-Yaus in the A-model interpretation. 
In the degree $5$ case, these two limits correspond to the large volume limit of the same Calabi-Yau. 
It was showed that our direct integration method is also useful to analyze the small moduli limit. 
It is interesting that we have the {\it hypersurface}-like period integral formulas 
after evaluating two of residue integrals, as noted in \eqref{hypersurface_like}: 
\beq
\Pi=\int \frac{C}{A-z^{1/2} B}. \nonum
\eeq
This is the common property for all the models analyzed in this paper, 
as seen in \eqref{13period}, \eqref{5period}, \eqref{7period} and \eqref{10period}. 
This may implies that our approach is powerful and effective for treating other Calabi-Yaus which are not analyzed so far due to their complexities.

The invariants predicted in this paper must be confirmed by the direct computation of the open Gromov-Witten invariants as done for the quintic in \cite{PaSoWa}. 
Such computations help us to fix the overall numerical constant and complete the study of the open mirror symmetry of the pfaffian Calabi-Yaus. 
For the pfaffian case, although the closed sector of the degree $14$ was already carried out in \cite{Tjotta}, the generalization to the open sector is not yet achieved. 
The first difficulty is that pfaffians are not complete intersections of sections of vector bundles but degeneracy loci of a certain morphism of vector bundles. 
The second difficulty is that we do not know the concrete constructions of special Lagrangian submanifolds in pfaffian Calabi-Yaus at present. 
This problem is rather serious since even in the case of complete intersections in a weighted projective space (e.g., the double quartic) we do not yet know their construction. 
Several these problems make the identification of A-model geometry and direct computation in the A-model side difficult. 
Furthermore, the concrete definition of A-model side Calabi-Yaus corresponding to the small moduli limits of $\check{X}_{13}$ and $\check{X}_{10}$ are still unknown. 
It is very interesting that by the mirror symmetry consideration, 
we can obtain real BPS invariants of such unknown Calabi-Yaus, although we do not know the concrete constructions of them at all. 
Our results indicate that there exist corresponding A-model geometries and Lagrangian branes with two discrete moduli.

Other interesting possibilities for further studies are the search for other branes with $\mathbb{Z}_k$-vacua ($k\neq 2$), as observed in \cite{Walcher}, and 
the analysis of the off-shell extension of the superpotential, discussed in \cite{JS1}. 
The analytic properties of the superpotentials in the entire open/closed moduli space are also worth studying. 
Moreover it is interesting to study the pfaffian Calabi-Yau manifolds with multiple moduli, since some of them may be useful to construct the phenomenological or cosmological models. 
They are beyond the scope of this paper and we leave it for future work.

We hope that the study of open mirror symmetry and the physics associated with the type II string theory compactified on {\it compact} Calabi-Yau with background D-brane geometry becomes more active and tractable by the appearance of this work. 
\\

\noindent{\bf Acknowledgements}
\\
M.S. would like to thank Yoshi Kondo and Johanna Knapp for helpful comments. 
H.S. is supported by a Grant-in-Aid for Scientific Research on Priority Area 
(Progress in Elementary Particle Physics of the  21st Century through Discoveries of Higgs Boson and Super- symmetry, Grant No. 16081201) 
provided by the Ministry of Education, Science, Sports and Culture, Japan. 

\appendix
\section{Tables of real BPS invariants}\label{invariants}
Here we list the tables of real BPS invariants at the tree level (half of disk invariants) up to an overall normalization. 
The possible overall numbers are discussed in section \ref{one-loop}. 
\begin{itemize}
\item \textbf{The degree $13$ (large moduli)} 
\end{itemize}
\begin{center}
\begin{tabular}{cr} \toprule
degree & $n_d^{(0,\text{real})}$ of $\check{X}_{13}$ \\ 
\midrule
1 & 7 \\ [-10pt] 
3 & 35 \\ [-10pt]  
5 & 2564 \\ [-10pt]  
7 & 270402 \\ [-10pt] 
9 & 32866812 \\ [-10pt] 
11 & 4517935956 \\ [-10pt] 
13 & 671632232977 \\ [-10pt] 
15 & 105623598511588 \\ [-10pt] 
17 & 17326521657555063 \\ [-10pt] 
19 & 2936975257157275841 \\ [-10pt] 
21 & 511056343067466075899 \\ [-10pt] 
23 & 90852752342596477299133 \\ [-10pt] 
25 & 16441821210636123211607972 \\ [-10pt] 
27 & 3020691730043208779353140253 \\ [-10pt] 
29 & 562169213445941154110358060705 \\ [-10pt] 
31 & 105798636416557155179367263056087 \\ [-10pt] 
33 & 20106435094248671076837246879637590 \\ [-10pt] 
35 & 3854160608672442333423080967857650538 \\ 
\bottomrule
\end{tabular}
\end{center}

\begin{itemize}
\item \textbf{The degree $13$ (small moduli)} 
\end{itemize}
\begin{center}
\begin{tabular}{cr} \toprule
degree & $n_d^{(0,\text{real})}$ of $\check{X}_{13}$ small moduli \\ 
\midrule
1 & 64a \\ [-10pt] 
3 & -47680a \\ [-10pt] 
5 & 2553150912a \\ [-10pt] 
7 & 34649296391104a \\ [-10pt] 
9 & 1471485784105332224a \\ [-10pt] 
11 & 49687346335244068056512a \\ [-10pt] 
13 & 1999219993905247049048775104a \\ [-10pt] 
15 & 82597293165445182789471202839616a \\ [-10pt] 
17 & 3586333115625143376407406077991120576a \\ [-10pt] 
19 & 160451134948695159327303909254189883152832a \\ [-10pt] 
21 & 7370533127928667910678110799417235303401680448a \\ [-10pt] 
23 & 345721234901527379438504092619993239987144802452416a \\ [-10pt] 
25 & 16503946288023845606947723040167603475881000397032091648a \\ [-10pt] 
27 & 799615568881349867687745900256952058894288795117507466374656a \\ [-10pt] 
29 & 39236388878593579256755137458766656519160016787082064331503911360a \\ 
\bottomrule
\end{tabular}
\end{center}
It is suggested that $a=0$, $13$, ..., by one-loop consideration.

\begin{itemize}
\item \textbf{The degree $5$} 
\end{itemize}
\begin{center}
\begin{tabular}{cr} \toprule
degree & the number of real BPS invariants in $X_{5}$ \\ 
\midrule
1 & 12a \\ [-10pt] 
3 & 556a \\ [-10pt] 
5 & 205552a \\ [-10pt] 
7 & 121112796a \\ [-10pt] 
9 & 86120101400a \\ [-10pt] 
11 & 69110942739196a \\ [-10pt] 
13 & 60075423135511800a \\ [-10pt] 
15 & 55275355154034182536a \\ [-10pt] 
17 & 53065617723187386623784a \\ [-10pt] 
19 & 52650031201378269968458204a \\ [-10pt] 
21 & 53628679366519667080820325372a \\ [-10pt] 
23 & 55810044042629530504437428769692a \\ [-10pt] 
25 & 59125681630282007996666603464358632a \\ [-10pt] 
27 & 63589410115281403817109787564072326184a \\ [-10pt] 
29 & 69277845189787603343632878895253381812556a \\ [-10pt] 
31 & 76322237467736882097065149698808737087298956a \\ [-10pt] 
33 & 84906997920284772853281422198585992071135299064a \\ [-10pt] 
35 & 95272751524203542192248811823850971401733034251936a \\ 
\bottomrule
\end{tabular}
\end{center}
It is suggested that $a=0$, $5$, ..., by one-loop consideration.

\begin{itemize}
\item \textbf{The degree $7$} 
\end{itemize}
\begin{center}
\begin{tabular}{cr} \toprule
degree & the number of real BPS invariants in $X_{7}$ \\ 
\midrule
1 & 10a \\ [-10pt] 
3 & 204a \\ [-10pt] 
5 & 43790a \\ [-10pt] 
7 & 14034754a \\ [-10pt] 
9 & 5377152402a \\ [-10pt] 
11 & 2324861044052a \\ [-10pt] 
13 & 1088221792755554a \\ [-10pt] 
15 & 539050098990610200a \\ [-10pt] 
17 & 278574103595914399154a \\ [-10pt] 
19 & 148775639332230303190266a \\ [-10pt] 
21 & 81568727403437332440801168a \\ [-10pt] 
23 & 45690577463967187842452075174a \\ [-10pt] 
25 & 26054107356717218372717843033150a \\ [-10pt] 
27 & 15082412998143462915292456125595662a \\ [-10pt] 
29 & 8844390475452660390527342087301322406a \\ [-10pt] 
31 & 5244624220491034544489908034567747495010a \\ [-10pt] 
33 & 3140502587424299686574001427880878557396696a \\ [-10pt] 
35 & 1896791544563405676821849252078336325607796030a \\ 
\bottomrule
\end{tabular}
\end{center}
It is suggested that $a=1$, $3$, ..., by one-loop consideration. 

\begin{itemize}
\item \textbf{The degree $10$ (large moduli)} 
\end{itemize}
\begin{center}
\begin{tabular}{cr} \toprule
degree & the number of real BPS invariants in $X_{10}$ \\ 
\midrule
1 & 8a \\ [-10pt] 
3 & 72a \\ [-10pt] 
5 & 7840a \\ [-10pt] 
7 & 1275496a \\ [-10pt] 
9 & 243115152a \\ [-10pt] 
11 & 52333685032a \\ [-10pt] 
13 & 12190638263120a \\ [-10pt] 
15 & 3004575113939760a \\ [-10pt] 
17 & 772507759536742768a \\ [-10pt] 
19 & 205250348613917160552a \\ [-10pt] 
21 & 55983141116972972765352a \\ [-10pt] 
23 & 15600497807514198642192616a \\ [-10pt] 
25 & 4425535856816745429812619760a \\ [-10pt] 
27 & 1274499263185215640279557261552a \\ [-10pt] 
29 & 371807289129697954226881629514120a \\ [-10pt] 
31 & 109685246417659296411075348428913288a \\ [-10pt] 
33 & 32675307792096894424635599988896855376a \\ [-10pt] 
35 & 9818152043216485859719280824348103702720a \\ 
\bottomrule
\end{tabular}
\end{center}
It is suggested that $a=0$, $2$, ..., by one-loop consideration.

\begin{itemize}
\item \textbf{The degree $10$ (small moduli)} 
\end{itemize}
\begin{center}
\begin{tabular}{cr} \toprule
degree & the number of real BPS invariants in $X_{10}$ \\ 
\midrule
1 & 16a \\ [-10pt] 
3 & 8048a \\ [-10pt] 
5 & 12744560a \\ [-10pt] 
7 & 34858414832a \\ [-10pt] 
9 & 120005634076032a \\ [-10pt] 
11 & 468860632792828784a \\ [-10pt] 
13 & 1988542693717684336240a \\ [-10pt] 
15 & 8935901455386424594255120a \\ [-10pt] 
17 & 41917867212967352197838814128a \\ [-10pt] 
19 & 203268668263311718810533932904048a \\ [-10pt] 
21 & 1012063908345796573491271043738275216a \\ [-10pt] 
23 & 5148565282099205877554243586387320508656a \\ [-10pt] 
25 & 26663710324297486380626275757344199971742720a \\ [-10pt] 
27 & 140183809674639050491540466368280185386449988480a \\ [-10pt] 
29 & 746566835578490949366745669972172742658825159266416a \\ [-10pt] 
31 & 4020464287771584663839492997988281211078691213564394480a \\ [-10pt] 
33 & 21862874813227538307930997168786349719887672243654857851120a \\ [-10pt] 
35 & 119910786858705051297319120366488737032816972266438414997855120a \\ 
\bottomrule
\end{tabular}
\end{center}
It is suggested that $a=0$, $10$, ..., by one-loop consideration.



\begin{thebibliography}{99}
\bibitem{Wa1}
J. Walcher, 
{\it Opening mirror symmetry on the quintic}, 
Commun. Math. Phys. {\bf 276} (2007), 671-689, 
arXiv:hep-th/0605162. 

\bibitem{MW}
D. R. Morrison and  J. Walcher, 
{\it D-branes and Normal Functions}, 
arXiv:0709.4028 [hep-th]. 

\bibitem{PaSoWa}
R. Pandharipande, J. Solomon and  J. Walcher, 
{\it Disk enumeration on the quintic $3$-fold}, 
J. Amer. Math. Soc. (2008), 
arXiv:math/0610901. 

\bibitem{BCOV1}
M. Bershadsky, S. Cecotti, H. Ooguri and C. Vafa, 
{\it Holomorphic anomalies in topological field theories}, 
Nucl. Phys. {\bf B405}, 279 (1993), 
arXiv:hep-th/9302103. 

\bibitem{BCOV2}
M. Bershadsky, S. Cecotti, H. Ooguri and C. Vafa, 
{\it Kodaira-Spencer theory of gravity and exact results for quantum string amplitudes}, 
Commun. Math. Phys. 165, 311(1994), 
arXiv:hep-th/9309140. 

\bibitem{Wa2}
J. Walcher, 
{\it Extended holomorphic anomaly and loop amplitudes in open topological string}, 
arXiv:0705.4098 [hep-th]. 

\bibitem{Wa3}
J. Walcher,
{\it Evidence for Tadpole Cancellation in the Topological String},
arXiv:0712.2775 [hep-th].

\bibitem{JS1}
H. Jockers and M. Soroush, 
{\it Effective superpotentials for compact D5-brane Calabi-Yau geometries}, 
Commun. Math. Phys. {\bf 290} (2009) 249-290, 
arXiv:0808.0761[hep-th]. 

\bibitem{LMW1}
W. Lerche, P. Mayr and N. Warner, 
{\it Holomorphic $N=1$ special geometry of open-closed type II strings}, 
arXiv:hep-th/0207259. 

\bibitem{LMW2}
W. Lerche, P. Mayr and N. Warner, 
{\it $N=1$ special geometry, mixed Hodge variations and toric geometry}, 
arXiv:hep-th/0208039. 

\bibitem{AHMM}
M. Alim, M. Hecht, P. Mayr and A. Mertens, 
{\it Mirror Symmetry for Toric Branes on Compact Hypersurfaces}, 
JHEP 0909:126,2009, 
arXiv:0901.2937 [hep-th]. 

\bibitem{AV}
M. Aganagic and C. Vafa, 
{\it Mirror symmetry, D-branes and counting holomorphic discs}, 
arXiv:hep-th/0012041. 

\bibitem{CDGP}
P. Candelas, X. C. de la Ossa,  P. S. Green and L. Parkes, 
{\it A pair of Calabi--Yau manifolds as an exactly soluble superconformal theory}, 
Nuclear Phys. {\bf B359} (1991), 21--74. 

\bibitem{JS2}
H. Jockers and M. Soroush, 
{\it Relative periods and open-string integer invariants for a compact Calabi-Yau hypersurface}, 
Nucl. Phys. {\bf B821} (2009) 535-552. 
arXiv:0904.4674 [hep-th]. 

\bibitem{KW1}
D. Krefl and J. Walcher,
{\it Real Mirror Symmetry for One-parameter Hypersurfaces},
JHEP 0809, 031 (2008), 
arXiv:0805.0792 [hep-th]. 

\bibitem{KS1}
J. Knapp and E. Scheidegger, 
{\it Towards Open String Mirror Symmetry for One-Parameter Calabi--Yau Hypersurfaces}, 
arXiv:0805.1013 [hep-th]. 

\bibitem{GHKK}
T. W. Grimm, T-W Ha, A. Klemm and D. Klevers, 
{\it The D5-brane effective action and superpotential in ${\cal N}=1$ compactifications}, 
Nucl. Phys. {\bf B816} (2009) 139-184,
arXiv:0811.2996 [hep-th]. 

\bibitem{Walcher}
J. Walcher, 
{\it Calculations for Mirror Symmetry with D-branes}, 
JHEP 0909:129,2009, 
arXiv:0904.4905 [hep-th]. 

\bibitem{AB}
M. Aganagic and C. Beem, 
{\it The Geometry of D-Brane Superpotentials}, 
arXiv:0909.2245 [hep-th]. 

\bibitem{worldsheet}
M. Baumgartl, I. Brunner and M. Soroush, 
{\it D-brane Superpotentials: Geometric and Worldsheet Approaches}, 
arXiv:1007.2447 [hep-th]. 

\bibitem{FNSS}
H. Fuji, S. Nakayama, M. Shimizu and H. Suzuki, 
{\it A Note on Computations of D-brane Superpotential}, 
to appear. 

\bibitem{mayr}
P. Mayr, 
{\it N = 1 mirror symmetry and open/closed string duality}, 
Adv. Theor. Math. Phys. 5 (2002) 213, 
arXiv:hep-th/0108229. 

\bibitem{AHJMMS}
M. Alim, M. Hecht, H. Jockers, P. Mayr, A. Mertens and M. Soroush,
{\it Hints for Off-Shell Mirror Symmetry in type II/F-theory Compactifications}, 
arXiv:0909.1842 [hep-th]. 

\bibitem{GHKK2}
T. W. Grimm, T.-W. Ha, A. Klemm and D. Klevers, 
{\it Computing Brane and Flux Superpotentials in F-theory Compactifications}, 
arXiv:0909.2025 [hep-th]. 

\bibitem{GHKK3}
T. W. Grimm, T.-W. Ha, A. Klemm and D. Klevers, 
{\it Five-Brane Superpotentials and Heterotic/F-theory Duality}, 
arXiv:0912.3250 [hep-th]. 

\bibitem{JMW}
H, Jockers, P, Mayr and J, Walcher, 
{\it On N=1 4d Effective Couplings for F-theory and Heterotic Vacua}, 
arXiv:0912.3265 [hep-th]. 

\bibitem{GKZ}
M. Alim, M. Hecht, H. Jockers, P. Mayr, A. Mertens and M. Soroush, 
{\it Type II/F-theory Superpotentials with Several Deformations and N=1 Mirror Symmetry}, 
arXiv:1010.0977 [hep-th]. 

\bibitem{data}
C. van Enckevort and D. van Straten, 
Electronic data base of Calabi-Yau equations, 
http://enriques.mathematik.uni-mainz.de/CYequations/ 

\bibitem{table}
G. Almkvist, C. van Enckevort, D. van Straten and Wadim Zudilin, 
{\it Tables of Calabi-Yau equations}, 
arXiv:math.AG/0507430. 

\bibitem{binomial}
G. Almkvist, 
{\it Some binomial identities related to Calabi-Yau differential equations}, 
arXiv:math/0703255 [math.CO]. 

\bibitem{Rodland}
E. A. R\o dland, 
{\it The Pfaffian Calabi-Yau, its Mirror, and their Link to the Grassmannian $Gr(2,7)$}, 
Composio Mathematica 122 (2000) no.2, 135-149, 
arXiv:math/9801092 [math.AG]. 

\bibitem{HoKo}
S. Hosono and Y. Konishi, 
{\it Higher genus Gromov-Witten invariants of the Grassmannian, and the Pfaffian Calabi-Yau threefolds}, 
Adv. Theor. Math. Phys. Vol.13 No.2 (2009), 463-495, 
arXiv:0704.2928 [math.AG]. 

\bibitem{Kanazawa}
A. Kanazawa, 
{\it On Pfaffian Calabi-Yau Varieties and Mirror Symmetry}, 
arXiv:1006.0223 [math.AG]. 

\bibitem{BBS}
K. Becker, M. Becker and A. Strominger, 
{\it Fivebranes, Membranes and Non-Perturbative String Theory}, 
Nucl. Phys. B456, 130 (1995), 
arXiv:hep-th/9507158. 

\bibitem{OOY}
H. Ooguri, Y. Oz and Z. Yin, 
{\it D-Branes on Calabi-Yau Spaces and Their Mirrors}, 
Nucl. Phys. B477, 407 (1996), 
arXiv:hep-th/9606112. 

\bibitem{OV}
H. Ooguri and C. Vafa, 
{\it Knot invariants and topological string}, 
Nucl.Phys. {\bf B577} (2000) 419-438, 
arXiv:hep-th/9912123. 

\bibitem{KKLM}
S. Kachru, S. H. Katz, A. E. Lawrence and J. McGreevy, 
{\it Open string instantons and superpotentials}, 
Phys. Rev. D 62, 026001 (2000),
arXiv:hep-th/9912151. 

\bibitem{Witten}
E. Witten, 
{\it Branes and the dynamics of {QCD}}, 
Nucl.\ Phys.\  B {\bf 507} (1997), 658--690, 
arXiv:hep-th/9706109. 

\bibitem{hCS}
E. Witten, 
{\it Chern-Simons Gauge Theory As A String Theory}, 
Prog. Math. 133, 637 (1995), 
arXiv:hep-th/9207094. 

\bibitem{Griffiths}
P. Griffiths, ed., 
{\it Topics in transcendental algebraic geometry}, 
in Proceedings of a seminar held at the Institute for Advanced Study, Princeton, NJ, 
during the academic year 1981/1982, 
Annals of Mathematics Studies {\bf 106}, 
Princeton University Press, Princeton, NJ, 1984. 

\bibitem{Griffiths2}
P. Griffiths, 
{\it On the periods of certain rational integrals: I}, 
Ann. Math. {\bf 90} (1969) 460. 

\bibitem{Green}
M. L. Green, 
{\it Infinitesimal methods in Hodge theory}, 
in Algebraic cycles and Hodge theory (Torino, 1993), 
Lecture Notes in Mathematics {\bf 1594}, Springer, Berlin, 1994, 1--92. 

\bibitem{LLY}
S. Li, B. H. Lian and S. T. Yau, 
{\it Picard-Fuchs Equations for Relative Periods and Abel-Jacobi Map for Calabi-Yau Hypersurfaces}, 
arXiv:0910.4215  [math.AG]. 

\bibitem{Tonoli}
F. Tonoli, 
{\it Construction of Calabi-Yau 3-folds in $\mathbb{P}_6$}, 
J. Alg. Geom. 13 (2004), 249-266. 

\bibitem{Boehm}
J. B\oe hm, 
{\it Mirror symmetry and tropical geometry}, 
arXiv:0708.4402 [math.AG]. 

\bibitem{Tjotta}
E. Tj\o tta, 
{\it Quantum cohomology of a Pfaffian Calabi-Yau variety: verifying mirror symmetry predictions}, 
Composio Mathematica 126 (2001), 78-89, 
arXiv:math/9906119 [math.AG]. 

\bibitem{HT}
K. Hori and D. Tong, 
{\it Aspects of Non-Abelian Gauge Dynamics in Two-Dimensional N=(2,2) Theories}, 
JHEP 0705:079,2007, 
arXiv:hep-th/0609032v2. 

\bibitem{BC}
L. Borisov and A. Caldararu, 
{\it The Pfaffian-Grassmannian derived equivalence}, 
arXiv:math/0608404 [math.AG]. 

\bibitem{Ku}
A. Kuznetsov, 
{\it Homological projective duality for Grassmannians of lines}, 
arXiv:math/0610957 [math.AG]. 

\end{thebibliography}
\end{document}